\documentclass[lettersize,journal]{IEEEtran}

\usepackage{tikz}
\usepackage{amsmath}

\usepackage{filecontents}

\AtBeginDocument{%
  }

\usepackage{amsmath,amssymb}

\usepackage{enumitem} %for leftmargin
\usepackage{balance}
\usepackage{mdframed}

\usepackage{booktabs}
\usepackage{url}
\usepackage{xurl}
\usepackage{xspace}
\usepackage{bbding}%\checkmark,\xsolidbrush
\usepackage{enumitem}% itemize size for leftmargin
\usepackage{pifont}% for \ding{}
\usepackage{amsmath}
\usepackage{pifont}
\usepackage{subfig} 
\usepackage{listings}
\usepackage{xcolor}
\usepackage{colortbl}
\usepackage{hyperref}
\usepackage{cleveref}
\definecolor{graya}{HTML}{e6e6e6}

\usepackage{multirow}
\usepackage{makecell}

\usepackage{array} 

\makeatletter
\newcommand\figcaption{\def\@captype{figure}\caption}
\newcommand\tabcaption{\def\@captype{table}\caption}
\makeatother
\usepackage[many]{tcolorbox}

\definecolor{darkgreen}{rgb}{0, 0.5, 0}

\newcommand{\cmt}[1]{\textcolor{red}{\textit{#1}}}

\newcommand{\dwj}[1]{\textcolor{black}{#1}}

\newcommand{\cs}[1]{\textcolor{black}{#1}}

\newcommand{\poccover}{$\text{Coverage}_{\text{PoC}}$\xspace}
\newcommand{\ourtool}{\textsc{Kait}\xspace}

\usepackage{listings}
\usepackage{xcolor}

\definecolor{codegreen}{rgb}{0,0.6,0} % can use it as well.

\lstdefinestyle{json}{
    basicstyle=\scriptsize\ttfamily,
    numbers=left,
    numberstyle=\scriptsize,
    stepnumber=1,
    numbersep=3pt,
    showstringspaces=false,
    breaklines=true,
    frame=lines,
    backgroundcolor=\color{white},
    stringstyle=\color{black},
    keywordstyle=\color{blue},
    numberstyle=\color{gray},
    commentstyle=\color{blue},
    morestring=[b]",
    morecomment=[l]{//},
    literate=
     *{0}{{{\color{orange}0}}}{1}
      {1}{{{\color{orange}1}}}{1}
      {2}{{{\color{orange}2}}}{1}
      {3}{{{\color{orange}3}}}{1}
      {4}{{{\color{orange}4}}}{1}
      {5}{{{\color{orange}5}}}{1}
      {6}{{{\color{orange}6}}}{1}
      {7}{{{\color{orange}7}}}{1}
      {8}{{{\color{orange}8}}}{1}
      {9}{{{\color{orange}9}}}{1}
}

\definecolor{keywordcolor}{rgb}{0,0,1} % Blue for keywords
\definecolor{stringcolor}{rgb}{0.5,0,0} % Dark red for strings

\lstdefinelanguage{HTTP}{
    morekeywords={POST, GET, Host, User-Agent, Accept, Accept-Language, Accept-Encoding, Content-Type, X-Requested-With, Content-Length, Origin, DNT, Connection, Referer, Cookie},
    sensitive=false, % keywords are not case-sensitive
    morestring=[b]", % strings are enclosed in double quotes
    morecomment=[s]{\{}{\}}, % comments are enclosed in curly braces (not typical in HTTP but just for illustration)
}

\begin{document}

% \title{Large-Scale Empirical Study of Vulnerability Proof-of-Concept in the Wild}
% \title{Which PoCs Are Usable? A Large-Scale Empirical Study of \\Vulnerability PoCs in the Wild}
%\title{Real-World Usability of Vulnerability Proof-of-Concepts:\\ Insights from A Large-Scale Study}
%\title{A Large-Scale Study on the Usability of Vulnerability\\Proof-of-Concepts in the Wild}
%\title{A Large-Scale Empirical Study on Vulnerability Proof-of-Concepts in the Wild}
%\title{Real-World Usability of Vulnerability Proof-of-Concepts: Insights from A Large-Scale Study}

\title{Real-World Usability of Vulnerability Proof-of-Concepts: A Comprehensive Study}

% \author{Anonymous Author(s)}
\author{Wenjing Dang, Kaixuan Li,~\IEEEmembership{Member,~IEEE}, Sen Chen,~\IEEEmembership{Member,~IEEE}, Zhenwei Zhuo, \\Lyuye Zhang,~\IEEEmembership{Member,~IEEE}, and Zheli Liu,~\IEEEmembership{Member,~IEEE}
\thanks{Wenjing Dang and Kaixuan Li contributed equally to this work.}
\thanks{Wenjing Dang and Zhenwei Zhuo are with the College of Intelligence and Computing, Tianjin University, China. Kaixuan Li and Lyuye Zhang are with the Nanyang Technological University, Singapore. Sen Chen (Corresponding author) and Zheli Liu are with the Nankai University, China. (email: dangwenjing@tju.edu.cn; kaixuan.li@ntu.edu.sg; senchen@nankai.edu.cn; 3021244351@tju.edu.cn; zh0004ye@e.ntu.edu.sg; liuzheli@nankai.edu.cn)}
}

% \renewcommand{\shortauthors}{Trovato et al.}

%%
%% The abstract is a short summary of the work to be presented in the
%% article.

% \keywords{Proof-of-Concept, Vulnerability, Empirical study}

\maketitle

\begin{abstract}
% Software security remains a critical concern in our increasingly digital world. 
% Vulnerability Proof-of-Concept (PoC) plays a crucial role in demonstrating the practical exploitability of vulnerabilities and serves as an essential tool for detecting, validating, and preparing responses to security threats. 
% However, there is a lack of comprehensive studies examining the current usability status of vulnerability PoCs in the wild, including the PoC prevalence of disclosed vulnerabilities, and the integrity and reproducibility of existing PoCs.

The Proof-of-Concept (PoC) for a vulnerability is crucial in validating its existence, mitigating false positives, and illustrating the severity of the security threat it poses. However, research on PoCs significantly lags behind studies focusing on vulnerability data. This discrepancy can be directly attributed to several challenges, including the dispersion of real-world PoCs across multiple platforms, the diversity in writing styles, and the difficulty associated with PoC reproduction.
%persistency, quality, and reproducibility.
To fill this gap, we conduct the first large-scale study on
%to understand the usability of 
PoCs in the wild, \cs{assessing their report availability, completeness, reproducibility.}
%To this end, we first performed a systematic literature review to identify available PoC platforms. 
Specifically, 1) to investigate PoC reports availability for CVE vulnerability,
we collected an extensive dataset of \cs{470,921 PoCs and their reports} from 13 platforms, representing the broadest collection of publicly available PoCs to date. 
\cs{2) To assess the completeness of PoC report at a fine-grained level, we proposed a component extraction method, which combines pattern-matching techniques with a fine-tuned BERT-NER model to extract 9 key components from PoC reports.} 
%We annotated 2,400 sampled PoC reports to create ground-truth data for fine-tuning the aspect extraction model. 
3) To evaluate the effectiveness of PoCs, we \cs{recruited} 8 participants to manually reproduce 150 sampled vulnerabilities \cs{with 32 vulnerability types based on PoC reports}, enabling an in-depth analysis of PoC reproducibility and the factors influencing it. 
%Our findings reveal 
%significant insights, including the lack of PoCs for 78.9\% of CVE vulnerabilities, 
%potential impact factors for \dwj{low persistency}, suboptimal quality of PoC reports, 
\cs{Our findings reveal that 78.9\% of CVE vulnerabilities lack \dwj{available} PoCs, and existing PoC reports typically miss about 30\% of the essential components required for effective vulnerability understanding and reproduction, with various reasons identified for the failure to reproduce vulnerabilities using available PoC reports.
%and challenges in reproducing vulnerabilities using available PoCs. 
%Building upon these findings, our study not only provides a systematic understanding of the current PoC research landscape but also highlights the critical need to address the usability of PoCs. 
Finally, we proposed actionable strategies for stakeholders to %improve and discuss future research directions to 
enhance the overall usability of vulnerability PoCs in strengthening software security.}
\end{abstract}

\section{Introduction}\label{sec:intro}
Software security remains a critical concern in our increasingly digital world.
%, where vulnerabilities can profoundly impact everything from individual privacy to global economic stability. 
%In 2023, approximately 28,000 new vulnerabilities were reported to Common Vulnerabilities and Exposures (CVE)~\cite{Vulnerab32:online}.
\cs{Every year, a substantial number of vulnerabilities are disclosed to the Common Vulnerabilities and Exposures (CVE).}
%, marking a significant increase from previous years~\cite{Vulnerab32:online}. 
The rise in the number of CVE entries highlights the growing challenge of cybersecurity as organizations continue to face an increasing number of threats. 

Proof-of-Concept (PoC) is a demonstration that showcases the practical feasibility of a particular vulnerability. It plays a critical role not only in traditional security scenarios, such as detecting, validating, and preparing responses to security threats but also in software supply chain security. 
Specifically, regarding vulnerability detection, PoCs can be used to verify the actual existence of vulnerabilities to reduce false positives of detection tools~\cite{krupp2018teether,perez2021smart}. \dwj{Meanwhile, PoCs can be used as seeds to assist in dynamic monitoring~\cite{kwon2021octopocs,you2017semfuzz}. 
For vulnerability remediation, PoCs help developers identify the location and path of the vulnerability, facilitating the remediation process~\cite{DBLP:conf/sp/VotipkaSRHM18} and verifying whether the vulnerability fix is effective. 
Additionally, PoC takes an important role to play in exploitability prediction~\cite{bozorgi2010beyond,wang2018revery,suciu2022expected}, migration attacks~\cite{dai2021facilitating, kwon2021octopocs, jiang2022evocatio}, etc.}
% 前面提了传统场景对于poc的迫切需要，实际上最新安全场景亦需要，比如供应链sca场景，需要基于基础poc能力做迁移，做适配，这些的基础都是建立在对poc的深刻理解之上。
Recently, PoCs have become invaluable in the software supply chain security domain. Specifically, PoCs can be adapted and applied to test vulnerabilities in software components and dependencies, enabling more comprehensive security analysis and hardening of the software supply chain~\cite{chen2023exploiting}. A deep understanding of PoC capabilities is critical for advancing software supply chain attack prevention and mitigation techniques. 
\dwj{Overall, PoC can be used not only to confirm the existence of vulnerabilities to mitigate false positives, but also plays an important role in a variety of subsequent research scenarios.}

However, significant research gaps persist in studies related to PoCs.
Currently, most existing studies focus on downstream security analysis, such as exploitability prediction~\cite{bozorgi2010beyond,wang2018revery,suciu2022expected} and migration attacks~\cite{dai2021facilitating, kwon2021octopocs, jiang2022evocatio}, which rely on PoCs as input for these tasks.
%Furthermore, while some studies on vulnerability \cs{or} bug  reports~\cite{mu2018understanding,chaparro2017detecting,bettenburg2008makes} have acknowledged the importance of PoCs, 
%\cs{Additionally, although some studies focus on vulnerability or bug reports and highlight the importance of PoCs~\cite{mu2018understanding,chaparro2017detecting,bettenburg2008makes}, none have specifically focused on treating PoCs as the primary subject of study, particularly in terms of assessing their real-world usability across large-scale PoC datasets.}
\dwj{Additionally, although some studies focus on vulnerability or bug reports and highlight the importance of PoCs~\cite{mu2018understanding,chaparro2017detecting,bettenburg2008makes}, they have not directly studied PoCs themselves.}
\cs{For example, Mu et al.~\cite{mu2018understanding} aimed to understand the vulnerability reproducibility based on CVE reports including description information and external references, which could be constituted by technical reports, blog/forum posts, or PoCs. They considered these references as the crowd-sourced vulnerability reports which contain detailed information for vulnerability reproduction. However, they did not focus on PoC and their reports written specifically for vulnerability reproduction. Therefore, a large-scale study on PoCs and their reports in the wild is still missing.}

To bridge these gaps, we conducted the first large-scale study toward understanding the usability of vulnerability PoCs in the wild, assessing \cs{their report availability, completeness, and reproducibility.} 
We are facing three challenges as follows: 
\textbf{\textit{C1: The Lack of Large-Scale PoC Dataset.}}
On the one hand, regarding academia, as previously discussed, \cs{no studies have been conducted on a large-scale dataset of PoCs and their reports in the wild.}
On the other hand, regarding the industry, although many PoC databases are proposed such as Exploit DB, CXSecurity, Metasploit, they are dispersed across numerous platforms (details in Section~\ref{sec:collectpoc}), resulting in the absence of a centralized and unified PoC data source. 
%Additionally, aggregating all available PoC reports from these sources requires substantial human effort to analyze and decode their data crawling mechanisms. 
These factors combined create significant barriers to collecting and using representative PoC datasets. 
\textbf{\textit{C2: Diverse PoC Report Formats.}}
Beyond the \cs{dispersed} databases, the formats of PoC reports vary widely, with many being unstructured text (details in Section~\ref{secbert}). 
This diversity complicates the task of conducting in-depth analyses of the \cs{completeness} of PoC reports, especially when attempting \cs{automated} analysis of a large-scale dataset collected from diverse sources. 
\textbf{\textit{C3: Non-trivial Manual Efforts for Reproduction.}}
The primary usage of PoC and its report is for the reproduction of vulnerabilities, a critical step in demonstrating their exploitability and assessing the overall \cs{effectiveness} of the PoCs. 
\cs{However, this reproduction process is labor-intensive and time-consuming, 
%and it relies heavily on the expertise of professionals skilled in programming and vulnerability exploitation, 
especially when dealing with a variety of vulnerability types.} 
%This necessity for expert involvement adds significant complexity to accurately evaluating the effectiveness and quality of PoC reports. 

To tackle the challenges, we systematically collected \cs{datasets of PoC and vulnerability} based on a systematic literature review. We spent over 600 man-hours collecting 470,921 PoC reports from 13 available databases and 234,982 CVE entries from CVE and National Vulnerability Database (NVD) as of \cs{July} 2024. 
\textit{\textbf{1)}} For all collected CVEs, we explored the \cs{availability} of PoCs from our collected large-scale dataset. 
We further discussed the impact factors of PoC with low \cs{availability}, including vulnerability types, program types, publish time, and severity of vulnerabilities, etc.
\textit{\textbf{2)}} For all collected PoC reports, we conducted a fine-grained \cs{completeness} assessment by defining \cs{key components} of PoC reports and developing a \cs{component extraction method by combining pattern-matching techniques and fine-tuned BERT-NER model.}
%\underline{K}ey \underline{a}spect \underline{i}dentification \underline{t}ool named \ourtool. 
\textit{\textbf{3)}} For vulnerabilities with PoC reports, we explored the reproducibility of PoCs and potential influencing factors by involving 8 security experts in manually reproducing 150 sampled vulnerabilities with 32 vulnerability types.
% , which took over 450 man-hours.

Our study unveils that 78.9\% of the CVE vulnerabilities lack available PoCs, with potential causes including incorrect omission of CVE IDs in PoC reports, 
%missing maintenance of databases, 
\cs{unavailable databases,}
and avoiding abuse of published attacks. 
Meanwhile, existing PoC reports generally lack 30\% key \cs{components} that are critical for vulnerability understanding and reproduction, especially for environment-setup data and trigger-related data. 
For vulnerability reproduction, only 28.0\% (42/150) of vulnerabilities can be reproduced directly. The primary causes are environmental deployment issues and incomplete PoC reports. Strategies such as PoC information fusion across sources, Large Language Model (LLM) assisted \cs{component} reasoning \& error diagnosis can effectively improve the reproduction ratio to 78.7\%. 

In summary, we made the following main contributions:
\begin{itemize}[leftmargin=20pt] 
    
    \item {\textbf{PoC Empirical Study.} By using the collected PoC dataset, we conducted the first large-scale study on PoC reports in the wild. This study investigates PoC reports across three dimensions, including \cs{availability, completeness, and reproducibility} to provide a comprehensive overview of the current usability status of PoCs.} 
    
    \item \textbf{New Tool Development.} We \cs{proposed a hybrid method} for \cs{identifying key components} of PoC reports, aiming to \cs{investigate the completeness of PoC reports} and enhance PoC \cs{completeness} by standardizing report formats. %\cs{The implementation of this method} will be open-sourced to support further research in this field. 
    
    \item {\textbf{Insights and Implications.} Based on the useful study results, we provided several insights that benefit multiple stakeholders, including PoC developers, maintainers, users, and researchers. We also discussed future research directions for the vulnerability PoC field to foster advancements in PoC-related research.}

    % \item \textbf{Available Data and Tool.} 
    % %\dwj{
    % %We systematically collected the largest PoC dataset to date, comprising 470,921 PoC reports. Each report is structurally annotated with key components, enabling aspect-level retrieval. Furthermore, }
    % To support the community and future research, we have published all details on \url{https://sites.google.com/view/poc-study}, and the tool on \url{https://anonymous.4open.science/r/KeyComponentExtraction/}.
    
    % \item \textbf{Open-sourced Dataset.} \dwj{We systematically constructed the largest PoC dataset to date, comprising 470,921 PoC reports. Each report is structurally annotated with key components, enabling aspect-level retrieval. 
    % % Notably, we have only published data that is publicly available without extra data.
    % % We now published all available PoCs by considering data re-distribution license usage sufficiently. 
    % % To support further reproduction and research by the community, we provide all relevant details and a dataset demo on \url{https://sites.google.com/view/poc-study/home}. 
    % Adhering to ethical standards and legal guidelines, we will publish PoCs compliant with data redistribution licenses.  Relevant details and a dataset demo are now available on \url{https://sites.google.com/view/poc-study/home}. 
    % }

\end{itemize}

\section{Background and Study Design}

\subsection{PoC Report and Its Key \cs{Components}}\label{sec:keyaspectsdefine}
\begin{figure}
  \centering
  \includegraphics[width=0.8\columnwidth]{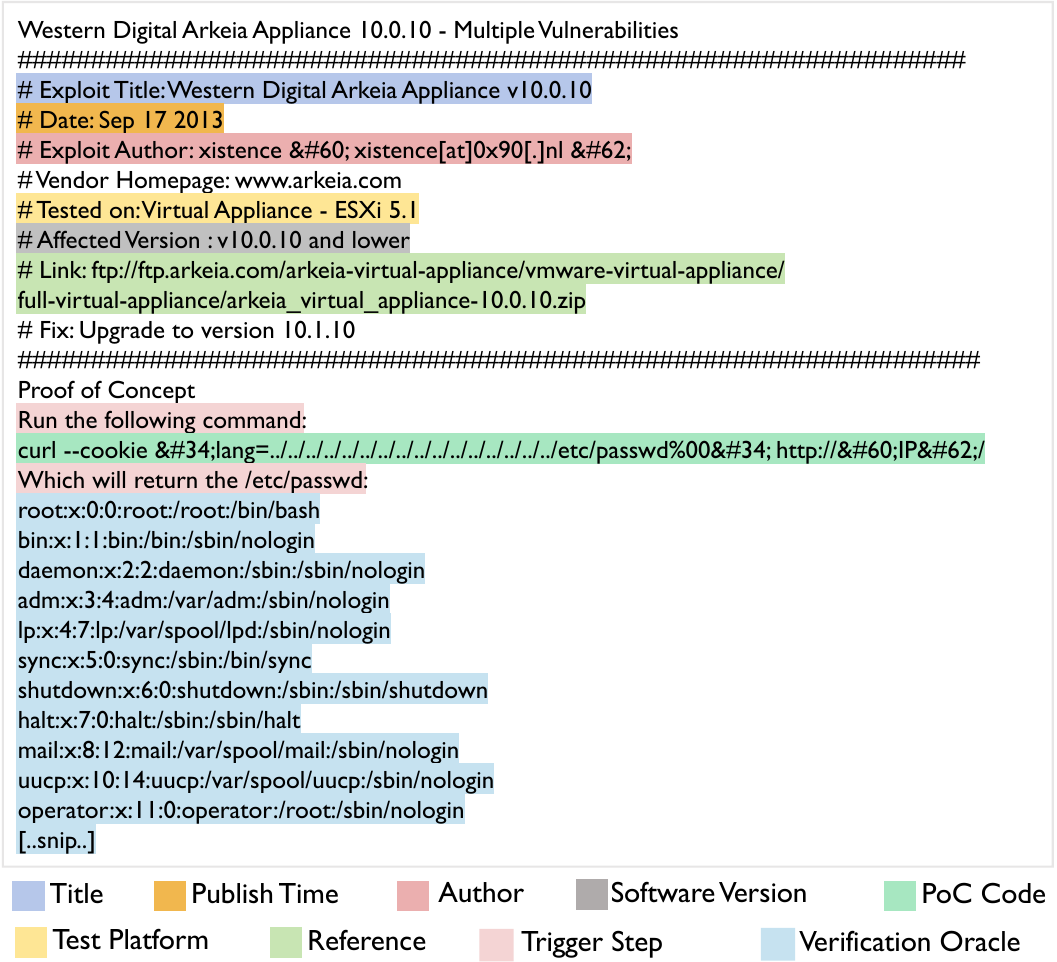}
  \caption{Example of a PoC report and its key \cs{components}.}
  \label{fig:poc-report}
\end{figure}

%A PoC report demonstrates the practical feasibility of a vulnerability. 
PoCs focus on verifying vulnerabilities for research and analysis purposes. 
% PoC not only confirms the existence of vulnerabilities but also plays a critical role in vulnerability remediation, detection, migration, and severity confirmation, etc.
PoC reports typically involve a small piece of code along with triggering steps to verify how the vulnerabilities can be triggered.
%It typically involves creating a small piece of code along with a triggering step to verify how the vulnerability can be triggered. 
% , and sometimes, it may only consist of a triggering step. 
%PoC can illustrate the severity of a vulnerability and help vendors understand the nature of the issue so they can develop corresponding patches.
%PoCs focus on verifying vulnerabilities for research and analysis purposes, while exploits are developed to achieve specific malicious outcomes. 
% A PoC is often considered a simplified version of an exploit. 
% Actually, besides the triggering step and PoC code (if exists), there are many other useful aspects provided in the wild. We take all these available data as \textbf{\textit{PoC report}}. 
%Following Mu et al.~\cite{mu2018understanding} and our in-depth investigation of a large number of PoC reports, 
% \revise
%{Drawing on prior research~\cite{mu2018understanding} and our in-depth investigation of a large number of PoC reports,}
\cs{Based on our in-depth investigation of a large number of PoC reports collected from various platforms,
we define 9 ``key components'' in three categories for the PoC report.}
% , which are important to PoC understanding and vulnerability reproduction. 
\textbf{Envirmonent Deployment Information}: \textit{Test Platform} and \textit{Software Version} serve as crucial foundations for setting up \cs{a vulnerability deployment environment.} 
\textbf{Vulnerability Trigger Information}: \textit{Trigger Step} describes specific steps for a vulnerability to be triggered. If the PoC is triggered through interactive operations, we could follow the steps in \textit{Trigger Step} to reproduce the vulnerability, and if it is triggered through code script, we would also need to follow the steps to run the \textit{PoC Code}. \textit{Verification Oracle} indicates that the vulnerability was successfully triggered. 
%The five aspects mentioned earlier are important to the success of vulnerability reproduction. 
\dwj{\textbf{Basic Information}: \textit{Title}, \textit{Publish Time}, \textit{Author}, and \textit{Reference} are important for understanding the life cycle PoC reports.}
%In summary, nine aspects are collectively named as \textbf{\textit{key components}}. 
{\Cref{fig:poc-report} shows an example of a PoC report, with different key components marked with distinct colors.} 

\subsection{Study Design}

\begin{figure*}[t]
  \centering
  \includegraphics[width=0.85\textwidth]{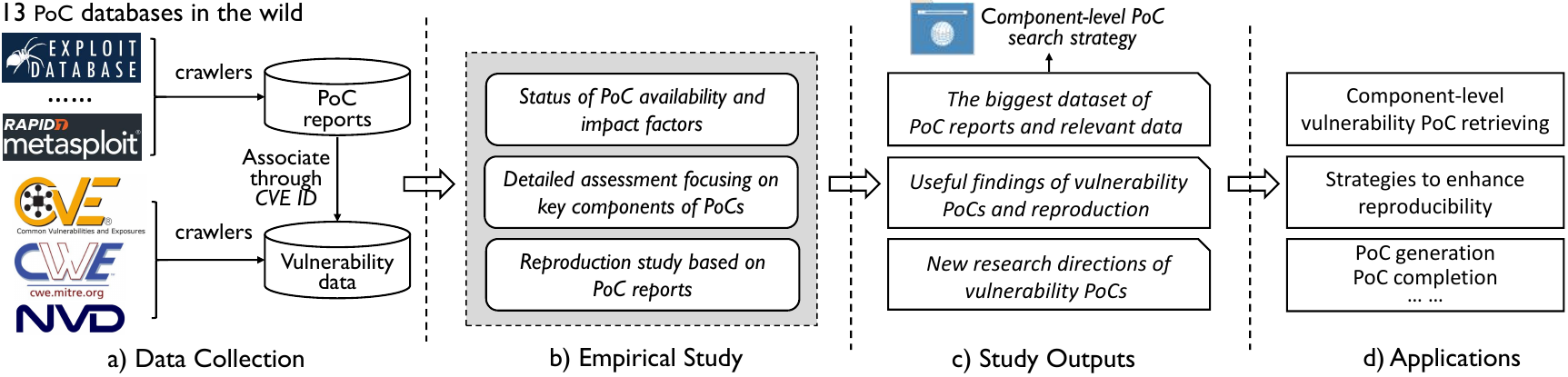}
  \caption{Overview of our study.}
  \label{fig:overview}
\end{figure*}

The goal of this paper is to conduct a large-scale empirical study on vulnerability PoCs in the wild, aiming to uncover valuable insights for this critical research area.
As shown in~\Cref{fig:overview}, we first performed a systematic collection for PoC and vulnerability data in~\Cref{DataCollection}. 
Regarding PoC data, we crawled all available PoC reports from diverse sources encompassing exploit databases (e.g., Exploit DB), vulnerability databases (e.g., Openwall), and code hosting platforms (e.g., GitHub). 
Regarding vulnerability data, we collected all disclosed vulnerabilities from vulnerability databases including CVE and NVD, complemented by Common Weakness Enumeration (CWE)'s categorization.

We designed our study by exploring 3 dimensions based on the large-scale datasets of PoC reports and vulnerability data.
\textbf{\textit{1) \cs{Availability}}} (details in Section~\ref{sec: poc-perspective}). For all collected CVEs, our goal is to explore whether we can obtain the corresponding PoC reports from our large-scale dataset. If not, what is the status of PoC \cs{availability} and the corresponding impact factors? \textbf{\textit{2) \cs{Completeness}}} (details in Section~\ref{secbert}). For all collected PoC reports, we aim to investigate how to conduct a fine-grained completeness assessment of them. \textbf{\textit{3) Reproducibility}} (details in Section~\ref{secrepro}). For vulnerabilities that have PoC reports, we plan to further observe the reproduction success rate of PoCs and the factors influencing it.

\section{Data Collection}\label{DataCollection}

\subsection{PoC Data Collection}\label{sec:collectpoc}

\cs{Based on our empirical investigations, two main methods can be used to collect PoC data. The first is through PoC databases, and the second is through external references associated with CVEs.}
%Based on our systematic observations, the first method has the following issues: XXX. Therefore, we opted for the second method.
\cs{For the first method,} we performed a systematic literature review (SLR) to identify available PoC databases used in top-tier Security and Software Engineering venues. Concretely, we used keywords like ``PoC'' and ``exploit'' to search papers published over the past two decades and finally obtained 32 research papers on vulnerability PoCs as an initial list. 
We then retained the papers in which the PoC databases they used were explicitly mentioned. After the search process and further manual screening, we retrieved 15 relevant papers that use PoC databases. 
Through the SLR, we initially obtained 11 PoC databases.% including Exploit Database, SecurityFocus, Packet Storm, etc.

For the second method, our systematic observations revealed that collecting PoCs from the \cs{external references} associated with CVE vulnerabilities is not feasible for the following reasons. First, \cs{the references could be constituted by technical reports, or blog/forum posts rather than a PoC. 
Even if the references contain PoCs, it remains challenging to identify which specific part of that reference constitutes the PoC. 
% Even if the references contain PoCs, it is challenging to identify whether a PoC exists or which specific part constitutes the PoC.
}
%the reference links often do not contain PoCs, and even when they do, it is challenging to identify whether a PoC exists or which specific part constitutes the PoC. 
For example, the existing repository maintained on GitHub~\cite{GitHubtr21:online} traverses CVE's reference and simply matches specific strings like ``PoC'' and ``Proof-of-Concept'' to identify PoCs. It would miss numerous PoC reports that do not explicitly contain such strings, like~\cite{phpfmv1721:online, IcingaWe49:online}. 
Meanwhile, it would incorrectly identify the presence of PoC because many websites only mention strings like ``PoC'' without publishing \cs{real} PoCs, like~\cite{Microsof93:online, SupportC73:online}. Specific cases are shown in. 
Instead, the NVD categorizes references using various tags. To focus specifically on PoCs, we identified and used references tagged as ``Exploit''.
%, which we refer to as ``exploit links''. 
This strategy helps avoid inadvertently collecting vulnerability references that do not contain PoCs. However, \cs{the references to the ``Exploit'' tag} span a wide range of primary domains (3,329 unique primary domains are associated with all CVE ``Exploit'' links), making it impractical to collect and analyze all of them. To mitigate this limitation, we crawled all of these links and traced them back to their primary sources of databases. 
Applying \textit{the Pareto Principle}~\cite{dunford2014pareto}, we found that the top 19 sources, each was referred more than 500 times by ``Exploit'' links, cover 75\% of the total links and expanding beyond these sources contribute minimally to our insights but consume unnecessary effort and resources. 
Hence, we obtained the top 19 sources as supplementary databases. 

The above process \cs{based on these two methods} yielded a total of 30 \cs{databases}, and after removing duplicates, we obtained 21 candidate sources. Based on this, we further found that 9 ones are unavailable due to various reasons. 
Specifically, 3 \cs{databases} have ceased maintenance, i.e., SecurityFocus, SecurityTracker, and archives.neohapsi; 5 \cs{databases}, including WPScan prohibited personal collection methods like web scraping due to their commercial purposes.
Ultimately, as shown in Table~\ref{tab:cate}, we identified 13 available PoC \cs{databases}, including Red Hat Bugzilla (abbr. Bugzilla), PHP Bug Tracking System (abbr. PHP::Bugs), Packet Storm (abbr. P. Storm), Exploit Database (Exploit DB), Seebug, CXSecurity, Chromium Monorail (abbr. Chromium), Openwall, GitHub, Hunter, Metasploit, Talos, and Vulnerability Lab (abbr. Vul. Lab). We strictly adhere to the data redistribution requirement by checking the data use agreement of each \cs{database}.

    \begin{table}
\renewcommand{\arraystretch}{1}
    \centering
    \caption{PoC databases used in existing works.
    % P. Storm means ``Packet Storm''. Exploit DB refers to ``Exploit Database''. Vul. Lab means ``Vulnerability Lab''.
    }
    \label{tab:cate}
    \scalebox{0.7}{\begin{tabular}{c|c|c|c|c|c|c|c|c|c|c|c|c|c}
    \hline
    \textbf{\rotatebox{90}{Papers}}                                     & \textbf{\rotatebox{90}{Bugzilla}} & \rotatebox{90}{\textbf{PHP::Bugs}} & \textbf{\rotatebox{90}{P. Storm}} & \textbf{\rotatebox{90}{Exploit DB}} & \textbf{\rotatebox{90}{Seebug}} & \textbf{\rotatebox{90}{CXSecurity}} & \textbf{\rotatebox{90}{Chromium}} & \textbf{\rotatebox{90}{Metasploit}} & \textbf{\rotatebox{90}{Talos}} & \textbf{\rotatebox{90}{Vul. Lab}} & \textbf{\rotatebox{90}{Openwall}} & \textbf{\rotatebox{90}{GitHub}} & \textbf{\rotatebox{90}{Hunter}} \\ \hline
    \rowcolor[HTML]{D9D9D9} 
    Bratus et al.~\cite{bratus2007hackers}   &                           &                    & \CheckmarkBold                     &                           &                 &                     &                   & \CheckmarkBold                   &                     &                &                   &                   &                 \\ \hline
    MetaSymploit~\cite{wang2013metasymploit} &                           &                    &                       &                           &                 &                     &                   & \CheckmarkBold                   &                     &                &                   &                   &                 \\ \hline
    \rowcolor[HTML]{D9D9D9} 
    Credal~\cite{xu2016credal}               &                           &                    &                       & \CheckmarkBold                         &                 &                     &                   &                    &                     &                &                   &                   &                 \\ \hline
    ZenIDS~\cite{hawkins2017zenids}          &                           &                    &                       &                           &                 & \CheckmarkBold                   &                   &                    &                     &                &                   &                   &                 \\ \hline
    \rowcolor[HTML]{D9D9D9} 
    A2c~\cite{kwon2017a2c}                   &                           &                    &                       & \CheckmarkBold                         &                 &                     &                   & \CheckmarkBold                   &                     &                &                   &                   &                 \\ \hline
    Xu et al.~\cite{xu2017postmortem}                        &                           &                    &                       & \CheckmarkBold                         &                 &                     &                   &                    &                     &                &                   &                   &                 \\ \hline
    \rowcolor[HTML]{D9D9D9} 
    Mu et al.~\cite{mu2018understanding}     &   \CheckmarkBold        &                    &                       & \CheckmarkBold                         &                 &                     &                   &                    &                     &                &     \CheckmarkBold     &                   &                 \\ \hline
    Gupta et al.~\cite{gupta2018hunting}                     &                           &                    &                       &                           &                 &                     &                   &                    &                     & \CheckmarkBold                 &                   &                   &                 \\ \hline
    \rowcolor[HTML]{D9D9D9} 
    Zhao et al.~\cite{zhao2020large}                         &                           &                    &                       &                           & \CheckmarkBold               &                     &                   &                    &                     &                &                   &                   &                 \\ \hline
    OCTOPOCS~\cite{kwon2021octopocs}         & \CheckmarkBold                         &                    &                       &                           &                 &                     &                   &                    & \CheckmarkBold                     &                &                   &                   &                 \\ \hline
    \rowcolor[HTML]{D9D9D9} 
    Gramatron~\cite{srivastava2021gramatron} &                           & \CheckmarkBold                  &                       &                           &                 &                     &                   &                    &                     &                &                   &                   &                 \\ \hline
    Paul et al.~\cite{paul2021security}                      &                           &                    &                       &                           &                 &                     & \CheckmarkBold                   &                    &                     &                &                   &                   &                 \\ \hline
    \rowcolor[HTML]{D9D9D9} 
    RAProducer~\cite{yuan2021raproducer}     &                           &                    &                       & \CheckmarkBold                         &                 &                     &                   &                    &                     &                &                   &                   &                 \\ \hline
    EE~\cite{suciu2022expected}              &                           &                    &                       & \CheckmarkBold                         &                 &                     &                   &                    &                     &                &                   &                   &                 \\ \hline
    \rowcolor[HTML]{D9D9D9} 
    Chen et al.~\cite{chen2023exploiting}                    &                           &                    &                       &                           &                 &                     &                   &                    & \CheckmarkBold                     &                &                   &                   &                 \\ \hline
    Our study               &  \CheckmarkBold                 &  \CheckmarkBold             &   \CheckmarkBold            &     \CheckmarkBold      &   \CheckmarkBold     &    \CheckmarkBold          &  \CheckmarkBold         & \CheckmarkBold                   & \CheckmarkBold              & \CheckmarkBold              & \CheckmarkBold                   & \CheckmarkBold                 & \CheckmarkBold                 \\ \hline
    
     \rowcolor[HTML]{D9D9D9} \rotatebox{90}{\# PoCs}                                                       & \rotatebox{90}{176,105}                   & \rotatebox{90}{80,477}             & \rotatebox{90}{52,109}               & \rotatebox{90}{45,784}                    & \rotatebox{90}{44,232}          & \rotatebox{90}{40,385}              & \rotatebox{90}{10,217}            & \rotatebox{90}{2,101}             & \rotatebox{90}{1,769}           & \rotatebox{90}{1,516}           & \rotatebox{90}{6,972}               & \rotatebox{90}{5,444}           & \rotatebox{90}{3,812}                      \\ \hline
    \end{tabular}}
\end{table}

% \smallskip
%\noindent{\textbf{Crawling process.}} 
%\subsubsection{Crawling Process}
To automate the acquisition of PoC reports from these 13 \cs{databases}, we organized a team of 5 collaborators to crawl the data synchronously. 
After an extensive process, we developed over 2,000 lines of Python crawler code.
% and it took a total of over 600 man-hours to complete the process. 
Through this effort, we obtained a \cs{large-scale} dataset of 470,921 PoC reports. \cs{It took a total of over 600 man-hours to complete the process.}

% \begin{figure}
%     \centering
%     \includegraphics[width=0.5\columnwidth]{img/form.pdf}
%       \caption{Distribution of triggering forms of PoCs along with programming languages.}
%       \label{fig:form}
% \end{figure}

% \smallskip
%\noindent{\textbf{Data sharing.}}
% \subsubsection{Data sharing}
% To facilitate future research, we have organized and standardized the format of the \textbf{470,921} PoC reports we collected and open-sourced them on \url{www.sctruster.com/cybersploit/}. 
% Meanwhile, we provide a more flexible and user-friendly searching strategy that allows users to search PoC reports based on various key components, rather than being limited to searching by report title or CVE ID alone. 
% To the best of our knowledge, this is \textbf{the first and largest PoC dataset} to date. 
% \revise{@wangyue, please enhance this presentation according to Xing's paper}

\subsection{Vulnerability Data Collection}
To observe the PoC availability of existing vulnerabilities, we collected disclosed vulnerability data from CVE, NVD, and CWE databases. We extracted all available vulnerability data, including {234,982} CVE entries from CVE and NVD as of {July 2024}.
Each CVE entity includes CVE IDs, vulnerability types, references, products, affected versions, etc. 

To further understand and analyze the characteristics of each vulnerability type and their correlation with PoCs, we gathered all available information on CWE, including descriptions, relations, applicable platforms, etc. Finally, we obtained 938 unique CWE IDs along with their information and all available ``CVE $\to$ CWE Mapping'' listed on the NVD. After removing entries that lacked CWE IDs, we identified 226,753 CWE IDs from 166,352 CVE entries.

\subsection{Statistical Summary of \cs{the Collected Data}}\label{secoverlap}

\begin{figure}
    \centering
    \includegraphics[width=0.65\columnwidth]{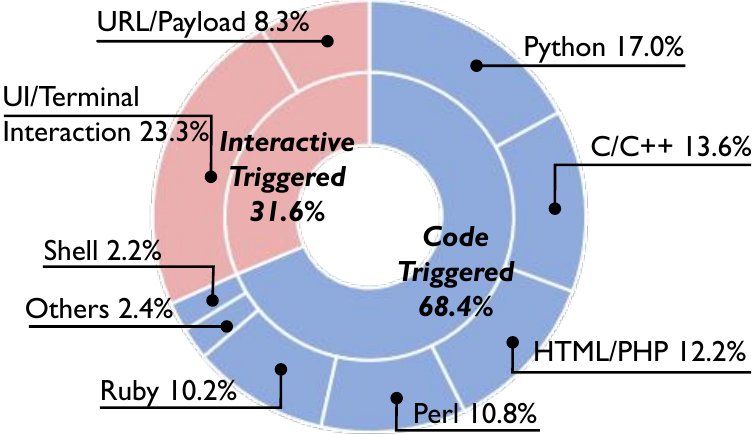}
      \caption{Distribution of triggering forms of PoCs along with programming languages.}
      \label{fig:form}
\end{figure}

%\cs{Stopped}
We show an overview of the current state of PoC reports on some basic characteristics.
% including {triggering forms} and corresponding CVE IDs. 
%\ding{182} 
\textbf{\textit{1)}} Triggering forms. \textit{Code Triggered} and \textit{Interactive Triggered} are the two types of triggering forms in PoC reports. 
\dwj{We define the features of 11 code languages and 2 forms of interaction triggers based on expert experience and in-depth research, and identify the departure forms of PoCs based on the feature matching approach.}
As shown in~\Cref{fig:form},
68.4\% of the PoC reports are in the form of Code Triggered, and the top 5 programming languages are Python, C/C++, HTML/PHP, Perl, and Ruby. The remaining 31.6\% are Interactive Triggered, typically providing a URL/Payload or detailing trigger steps involving UI/terminal interaction. 
\textbf{\textit{2)}} CVE ID presence. \Cref{subfig:cve-distribution} illustrates that only 18.5\% (87,251/470,921) of the PoC reports in our dataset carry a CVE ID, covering 49,667 unique CVE IDs. Many PoC reports correspond to the same CVE ID. 
\Cref{subfig:poc-distribution} shows that, while the three primary sources with the most PoC reports containing CVE IDs collectively account for 64,238 PoC reports with CVE IDs, they cover only 33,218 unique ones. Our detailed analysis revealed that 4,434 PoC reports share the same CVE IDs across all three \cs{databases}. 
\textbf{\textit{3)}} PoC duplication. {We observed a notable occurrence of PoC duplication across CVE vulnerabilities.} 
Specifically, among the 49,667 CVE vulnerabilities with associated PoCs, 51.7\% (25,689/49,667) correspond one-to-one with a single PoC. In contrast, a significant number demonstrate a one-to-many relationship: 44.9\% (22,314/49,667) are associated with between 2 and 5 PoCs, 2.7\% (1,357/49,667) with between 6 and 10 PoCs, and 0.6\% (307/49,667) CVEs have more than 10 corresponding PoCs. 
This duplication is partly due to repeated submissions of PoCs and partly because a single vulnerability may have multiple potential exploitation vectors~\cite{wang2013automatic}. 
To assess the extent of PoC duplication, we computed the percentage of duplicate PoC commits at the textual level using the longest common subsequence (LCS) algorithm~\cite{Longestc98:online}. Among the 19,283 CVE vulnerabilities with multiple textual-form PoCs, our analysis revealed that 15.8\% (3,047/19,283) involved duplicate submissions.

\section{Empirical Study}
% \cs{Firstly, we should state purpose for each part}
% \begin{figure}
% 	\centering
% 	\subfloat[Distribution of PoC reports with and without CVE IDs for all CVEs.]{\includegraphics[width=.24\columnwidth]{img/cvecoverage2.pdf}\label{subfig:cve-distribution}}\hspace{3pt}
% 	\subfloat[\footnotesize Distribution of PoC reports with and without CVE IDs for all PoC reports across all 13 different types of sources.]
%  {\includegraphics[width=.65\columnwidth]{img/cvecoverage1.pdf}\label{subfig:poc-distribution}}
% 	\caption{Distribution of PoC reports by CVE IDs.}
%     \label{fig:distribution}
% \end{figure}

%\subsection{PoC Study from Different Perspectives} \label{sec: poc-perspective}
% \subsection{Status of PoC Deficiencies and \\Impact Factors} \label{sec: poc-perspective}
\subsection{\cs{PoC Availability and Impact Factors}} \label{sec: poc-perspective}
Previous research has highlighted a substantial lack of PoCs~\cite{wang2018revery, mu2018understanding}. 
To quantify the availability, we first measure the percentage of CVE vulnerabilities that have corresponding PoC reports and explore the potential underlying reasons in~\Cref{secPoCCoverage}. Secondly, we investigate the differences in PoC coverage among various vulnerability types, as previous research related to PoC generation~\cite{brumley2008automatic, you2017semfuzz, chenefficient, cao2023oddfuzz, yang20231dfuzz} has often focused on specific vulnerability types, detailed in~\Cref{secvultypes}. 
{Furthermore, we also measured the impact of release time, the severity of vulnerabilities, and program types on PoC availability, discussed in~\Cref{secotherfactors}
% details are shown on~\cite{impactfactors} due to page limit.}\revise{Need results and findings}

% In~\Cref{PoCCoverage} (\textit{PoC Coverage on CVEs}), we investigated the probability of CVE vulnerabilities for which corresponding PoCs can be found, a metric we refer to as ``\poccover'', and analyzed the potential factors contributing to its absence. 
% In~\Cref{vultype} (\textit{Relations between PoC and Vulnerability Types}), we delved into the relations between the absence of PoCs and the types of vulnerabilities.

\begin{figure}
	\centering
	\subfloat[\footnotesize Distribution of CVE with and without PoC reports for all CVEs.]{\includegraphics[width=.33\columnwidth]{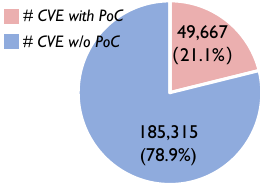}\label{subfig:cve-distribution}}
        \hspace{0pt}
	\subfloat[\footnotesize Overlap of CVE IDs among PoC \cs{databases}.]
{\includegraphics[width=.35\columnwidth]{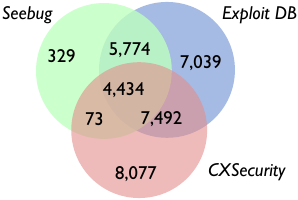}\label{subfig:poc-distribution}}
	\caption{Distribution of PoC reports by CVE IDs.}
    \label{fig:distribution}
\end{figure}

%\subsubsection{PoC Coverage on CVEs}\label{secPoCCoverage}
% \subsubsection{Status of CVEs with PoCs}\label{secPoCCoverage}
\subsubsection{\cs{Assessment of PoC Availability for CVE}}\label{secPoCCoverage}
% We utilize CVE entries in the CVE database as the baseline dataset \cmt{baseline dataset is precise? I am unsure.} 

% As of January 9, 2024, the CVE database has disclosed a total of 234,982 CVE IDs.
%CVE ID is one of the most popular identifiers used to uniquely record and retrieve vulnerabilities~\cite{wu2020cve} and is widely adopted by PoC databases, 
CVE ID is the identifier used to uniquely record and retrieve vulnerabilities and is widely adopted by \cs{databases}, 
which is required when submitting PoCs in 11 of the 13 \cs{databases} excluding GitHub and Openwall. 
Hence, to quantify how many existing vulnerabilities lack PoCs, we measured the percentage of all CVE vulnerabilities that have corresponding PoC reports in our dataset.
We define and calculate the PoC coverage as $\text{Coverage}_{\text{PoC}} = \frac{\# \text{ CVEs with PoC}}{\# \text{ CVEs}}$ by extracting the CVE IDs from the collected PoC reports. 
Based on our measurements, out of the 234,982 CVE entries, only 49,667 of them could find corresponding PoC reports in our dataset. 
\poccover is only 21.1\%, indicating about 80\% CVE entries missing the corresponding PoC reports.
% \revise{Plz, we need to describe it from CVE side instead of PoC side, if not , it is same to the last section!}
To this end, we attempt to explore several potential factors from different perspectives.

{\noindent{\textbf{\textit{Negligence in submission}}}:} 
%The inclusion of CVE IDs in PoC reports is typically not mandatory, 
\cs{The submission form of PoC databases is typically not manadatory,}
such as PoCs in \cs{GitHub} and Openwall, and the lack of a checksum mechanism often results in frequent omissions or errors during the submission process.
% The submission of CVE ID in PoC reports is often not mandatory, 
% The omission of CVE IDs may be attributed to negligence during the submission process of the PoC report. 
For example, a PoC report collected from CXSecurity~\cite{PHP5212544:online} originally carried a CWE ID within the field requiring the submission of a CVE ID. It was confirmed as a PoC report pointing to CVE-2009-4143. 

{\noindent{\textbf{\textit{Synchronization delays}}}:}
%\ding{182} \textit{Synchronization delays.} 
PoC reports may be published before their corresponding CVE IDs are assigned, but they were not updated after the CVE ID was obtained. An illustrative example involves the PoC report~\cite{HMAVPN5352:online} released on CXSecurity on February 22, 2022. Its associated CVE-2022-26634 was not published until three months later, resulting in PoC reports not being updated in a timely manner. 

%\ding{183} \textit{Negligence in submission.} 

{\noindent{\textbf{\textit{Non-coverage of disclosed vulnerabilities}}}:}
%\ding{184} \textit{Non-coverage of disclosed vulnerabilities.} 
Certain vulnerabilities detailed in PoC reports may never receive CVE IDs. Reasons can include a lack of recognition of the vulnerability's significance, absence of vendor confirmation, or oversights in the CVE assignment procedure, and some vendors maintain their own vulnerability databases and do not tend to apply for CVE IDs. 
This finding underscores the challenges faced by security researchers in accurately identifying PoC data due to missing or incorrect CVE IDs, which can significantly impact vulnerability management and response efforts. 

\noindent{\textbf{\textit{Lack of maintenance of PoC data}}}: 
As mentioned in Section~\ref{sec:collectpoc}, many 
%tagged ``Exploit'' links 
\cs{references tagged ``Exploit''}
on the NVD point to websites that have ceased maintenance, indicating the potential inaccessibility of numerous PoC reports. 
For example, our analysis revealed that 11,222 \cs{references} were directed to SecurityFocus, suggesting that over 10k PoC reports are now inaccessible due to the website maintenance discontinuation. 
Despite this, NVD continues to include these outdated 
%exploit URLs within its reference links,
\cs{references}
which significantly compromises the quality and accessibility of \cs{PoC} data. 
%This shortfall not only restricts exploitation technique information, hindering security research and vulnerability assessments but also affects cybersecurity training and policy development. 

\noindent{\textbf{\textit{Avoiding abuse of published attacks}}}:
There are many official channels for reporting vulnerabilities, such as CVE and NVD.
%CNVD~\cite{CNVD78:online}, and CNNVD~\cite{CNNVD14:online}. 
In contrast, PoC reports submitted to security vendors may be chosen to be withheld from disclosure. This would be due to security considerations, as some vulnerabilities cannot be fixed promptly and may still pose significant risks.

% \dwj{A significant number of vulnerabilities have been disclosed but not fully patched~\cite{zhang2023mitigating}. Releasing the PoC too quickly can pose a substantial security risk. For example, a PoC exploit for a vulnerability in Chrome browsers was released before the patch was fully deployed to users, allowing attackers to exploit the vulnerability in environments where the sandbox was disabled, leading to universal cross-site scripting attacks and unauthorized access to user accounts~\cite{PoCExplo37:online}. Therefore, for security reasons, some PoC reports may be deliberately withheld from the public to prevent the exploitation of these vulnerabilities and mitigate potential harm.} 
% For security reasons, some PoC reports\dwjcmt{add cite} may be deliberately withheld from the public to prevent the exploitation of vulnerabilities and potential harm. }
% \revise{Such description is too superficial! Plz start by describing the concrete phenomenon and then make a conclusion}

\begin{tcolorbox}[size=title,opacityfill=0.1,breakable]
\noindent \textit{\textbf{Remark:}} {{\textit{The lack of PoCs is severe, with only 21.1\% PoC coverage for CVE. 
{Potential causes include {negligence in submission, synchronization delay, non-coverage of disclosed vulnerabilities, lack of maintenance of PoC data, and avoiding abuse of published attacks.}}
% We explore several potential reasons for the missing PoCs and propose several ways to improve PoC coverage: 
% We explore several potential reasons for the missing PoCs and propose several ways to improve PoC coverage: 
% \ding{182} Paying attention to the link between vulnerability data and PoC data to avoid omitting the corresponding vulnerability information when submitting PoCs. \ding{183} Paying attention to the maintenance of the open-source PoC database and updating the failure data in a timely manner. \cmt{improve it a little.}
}}}
\end{tcolorbox}

% \begin{tcolorbox}[size=title,colframe=black,colback=white,breakable]
% \noindent \textbf{Remark:}{
% \textit{
% The low PoC coverage for CVE vulnerabilities significantly hampers security professionals' ability to deeply understand and verify these vulnerabilities. We recommend several measures to enhance PoC coverage and availability: completing PoC reports that lack CVE IDs with the appropriate identifiers, improving the open-source ecosystem for PoCs to avoid the sudden shutdown of large databases, and allocating more attention to CVE vulnerabilities without PoCs to reduce redundancy in PoC data.
% }
% }
% \end{tcolorbox}

\begin{table*}
%\footnotesize
\centering
\caption{{$\text{Coverage}_{PoC}$ in the CWE Top 25 list, highlighting the highest/lowest five CWEs regarding $\text{Coverage}_{PoC}$. The arrows indicate the deviation in \poccover for a vulnerability type compared to the average coverage rate.  
\textcolor{red}{$\downarrow$}/\textcolor{green}{$\uparrow$} indicate lower/higher than the overall average. 
% ``Buffer Overflow'' represents ``Improper Restriction Operations within the Bounds of a Memory Buffer''. 
}
} 
\label{tab:CWE25_PoC}
\scalebox{0.8}{\begin{tabular}{@{}cr|cr@{}}
\hline
\multicolumn{2}{c|}{{Last-5 CWE IDs}} & \multicolumn{2}{c}{Top-5 CWE IDs} \\ \hline
CWE ID         & $\text{Coverage}_{PoC}$               & CWE ID       & $\text{Coverage}_{PoC}$               \\ \hline
CWE-125 (Out-of-bounds Read)        & 10.3\% (\textcolor{red}{$\downarrow$} 9.8\%)       & CWE-94 (Code Injection)       & 43.8\% (\textcolor{green}{$\uparrow$} 23.7\%)      \\
CWE-77 (Command Injection)         & 10.7\% (\textcolor{red}{$\downarrow$} 9.4\%)       & CWE-89 (SQL Injection)       & 38.7\% (\textcolor{green}{$\uparrow$} 18.6\%)      \\
CWE-476 (NULL Pointer Dereference)        & 13.1\% (\textcolor{red}{$\downarrow$} 7.0\%)        & CWE-22 (Path Traversal)       & 32.8\% (\textcolor{green}{$\uparrow$} 12.7\%)      \\
CWE-269 (Improper Privilege Management)        & 13.3\% (\textcolor{red}{$\downarrow$} 6.9\%)       & CWE-416 (Use After Free)      & 28.1\% (\textcolor{green}{$\uparrow$} 8.0\%)       \\
CWE-863 (Incorrect Authorization)       & 14.2\% (\textcolor{red}{$\downarrow$} 5.9\%)       & CWE-119 (Buffer Overflow)      & 27.8\% (\textcolor{green}{$\uparrow$} 7.7\%)        \\ \hline
\end{tabular}}
\end{table*}

\subsubsection{Relations between PoC Coverage and Vulnerability Types}\label{secvultypes}
% \noindent{\textbf{\textit{Relations between PoC and Vulnerability Types:}}}
Previous research focusing on PoC generation~\cite{brumley2008automatic, you2017semfuzz, chenefficient, cao2023oddfuzz, yang20231dfuzz} has often concentrated on specific vulnerability types, suggesting a potential correlation between PoC availability and these types. 
Therefore, we measured \poccover across various vulnerability types. 
Using the aforementioned ``CVE $\to$ CWE Mapping'', we categorize PoC reports and their corresponding CVE IDs based on the associated CWE IDs. 
Given the impracticality of discussing each vulnerability type in detail, we focus on the latest CWE Top 25 list~\cite{CWECWETo97:online}.

The measurements indicate significant differences in \poccover across vulnerability types. 
\Cref{tab:CWE25_PoC} illustrates the Top-5 and Last-5 CWE IDs regarding \poccover, which ranged from as low as 10.3\% to as high as 43.8\%, and highlights the discrepancy between the \poccover for each type and the average \poccover. 
% This variation suggests that while some vulnerabilities are scrutinized, others may not receive as much attention. 
% This may be due to different perceptions of risk or ease of exploitation.
To explore the reasons behind the \cs{difference} in \poccover across vulnerability types, we further examined their characteristics and discussed the types of vulnerabilities with high \poccover and those with low \poccover, respectively. 

First, we start by discussing the top-5 vulnerability types with the highest \poccover, and we have identified several common characteristics: 
\textbf{\textit{1)}} \textit{Ease of exploitation.} These vulnerabilities typically do not require sophisticated techniques to exploit, particularly in applications that lack robust input validation and sanitization mechanisms. 
{Especially, CWE-94 (code injection) is relatively easier than CWE-77 (OS command injection) in that an attacker is only limited by the functionality of the injected language itself although both belong to injection vulnerabilities~\cite{CodeInje6:online}.}
\textbf{\textit{2)}} \textit{High impact.} Successful exploitation can lead to significant consequences, such as CWE-94 (Code Injection) typically leads to serious remote code execution problems, making it of high attention to researchers and attackers~\cite{Improper86:online}. 
\textbf{\textit{3)}} \textit{A wealth of extant research.} There exists a rich ecosystem of knowledge, tools, and PoCs for these vulnerabilities. This abundance facilitates the development and dissemination of PoCs by both researchers and attackers. For instance, numerous studies focus on the generation of PoC for CWE-89 (SQL Injection)~\cite{martin2008automatic, kieyzun2009automatic, alhuzali2016chainsaw, alhuzali2018navex}. 
\textbf{\textit{4)}} {\textit{Versatility of exploit patterns.}} 
PoCs targeting specific vulnerability types often exhibit obvious exploit patterns, especially in the trigger steps. 
For example, CWE-89-related PoCs typically involve similar steps and commonly used payloads (\texttt{' OR '1'='1}, \texttt{' OR '1'='1' --}, \texttt{'; DROP TABLE users; --}, etc.) like bypassing authentication, commenting out code, or executing stacked queries. 
This recurring pattern suggests the feasibility of extracting and developing generalized PoC templates for such vulnerabilities. Integration with existing SQL injection payload databases could simplify this process~\cite
{swisskyr30:online}. These templates could significantly automate and enhance the efficiency of PoC implementation for vulnerabilities characterized by recognizable exploit patterns.  %Moreover, the ability to outline clear PoC templates also contributes to the observed higher \poccover for certain types of vulnerabilities~\cite{martin2008automatic,sqlmappr75:online}. 
% The first payload attempts to bypass login authentication by changing the condition to an expression that is always true, the second one comments out subsequent SQL code using the comment symbol ``(--)'' and the last one uses stacked queries to execute multiple commands to delete tables from the database. 
% This indicates the possibility of extracting and developing a general PoC template for specific vulnerability types, which would greatly improve the efficiency of PoC implementation. 
% These factors collectively contribute to the prominence among vulnerability types with higher PoC coverage. 

% \begin{lstlisting}[language=HTTP, caption={Commonly used payload for CWE-89.}, label={lst:CWE89},float=t]
% ' OR '1'='1, 
% ' OR '1'='1' --, 
% '; DROP TABLE users; --
% \end{lstlisting} 

Furthermore, we discuss the last-5 vulnerability types with the lowest \poccover. We have also identified several possible reasons for this phenomenon. 
\textbf{\textit{1)}} \textit{Configuration-specific exploitation.} 
PoCs of CWE-269 (Improper Privilege Management) are heavily reliant on precise system settings and permissions. 
These configuration dependencies necessitate the creation of highly customized PoCs, which are not easily generalized across different environments.
% There is a strong dependency on the environment for exploitation. For example, CWE-269 (Improper Privilege Management) usually involves intricate system configurations and permission settings, which leads to the PoC specific to a particular setup or system design, making it difficult to write in bulk. 
{For instance, CVE-2023-1762 and CVE-2023-2240 are both classified as CWE-269 with similar vulnerability profiles. However, their PoCs differ significantly due to their reliance on specific system calls.
% , as discussed in \Cref{secappcompare}.
} 
\textbf{\textit{2)}} \textit{The uncertainty of vulnerability characteristics.} Vulnerabilities like CWE-125 (Out-of-bounds Read) and CWE-476 (Null Pointer Dereference) often do not directly lead to system crashes or data corruption. Specifically, they may result in information leaks and serve as parts of more complex attack chains while isolated vulnerabilities may not pose an immediate threat. 
Because of the uncertainty of the trigger step and resulting consequences, writing PoCs for these vulnerabilities is non-trivial. 
These factors combine to influence these types, giving them a lower \poccover. 

\begin{tcolorbox}[size=title,opacityfill=0.1,breakable]
\noindent \textit{\textbf{Remark:}} \textit{{
% There is a distinct difference in \poccover across vulnerability types, 
\poccover varies across vulnerability types,
{with the highest at 43.8\% and the lowest at 10.3\%.}
Types with high \poccover have the common characteristics of ease of exploitation, high impact, a wealth of extant research, and versatility of exploit patterns, such as CWE-94 and CWE-89. While those with low \poccover can be characterized by configuration-specific exploitation, such as CWE-269, and the uncertainty of vulnerability characteristics, such as CWE-125 and CWE-476.
% There are still many vulnerabilities for which PoCs are very seriously missing and PoC writing for these vulnerability types should be given high priority despite the difficulty of developing them. 
% We call on security personnel to prioritize writing PoCs for underappreciated vulnerability types shown in Table~\ref{tab:CWE25_PoC} to facilitate vulnerability understanding and remediation. 
}}
\end{tcolorbox}

%\subsubsection{Exploration on Other Potential Factors}\label{secotherfactors}
\subsubsection{Relations between PoC Coverage and other Aspects}\label{secotherfactors}

\begin{figure}
  \centering
  \includegraphics[width=0.9\columnwidth]{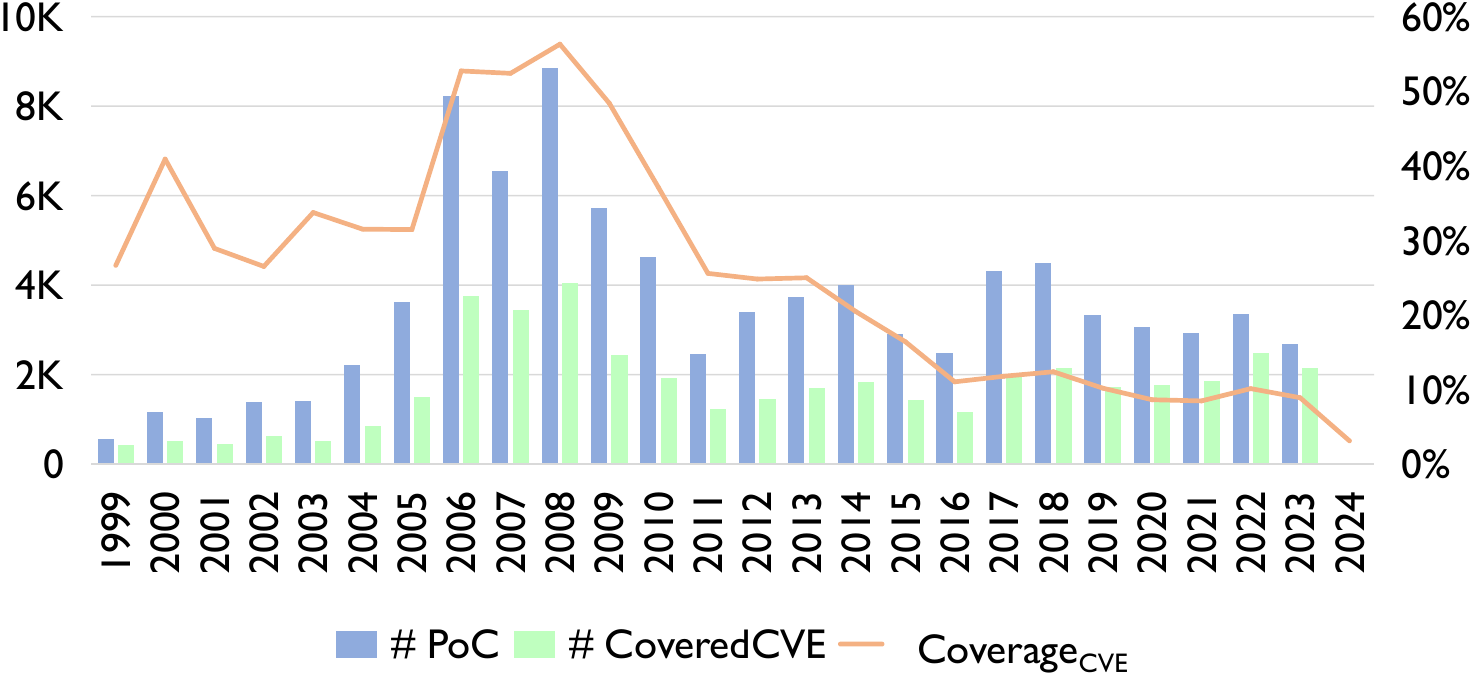}
  \caption{Relation between PoC coverage and release time.}
  \label{fig:time.pdf}
\end{figure}

In this section, we further measured the potential relations of \textit{release time, the severity of vulnerabilities, and program types} on PoC availability. 
\textbf{\textit{1)}} Release time. \dwj{As shown in~\Cref{fig:time.pdf}, the trend in PoC coverage over the years concerning CVE publish time can be summarized into three periods: \textit{Period of a plateau (1999-2005)}, \textit{Period of growth (2005-2008)}, and \textit{Period of contrarianism (2008-2024)}. }
\textbf{\textit{2)}} Vulnerablity severity. 
{
The Common Vulnerability Scoring System (CVSS)~\cite{CommonVu43:online} serves as an open industry standard to assess vulnerability severity and guide the prioritization of response measures. 
To provide a comprehensive analysis, we examined both the CVSS v3.1 scoring system (including four levels: LOW, MEDIUM, HIGH, and CRITICAL) and the CVSS v2 scoring system (including three levels: LOW, MEDIUM, and HIGH). Our in-depth analysis indicates that higher CVSS levels correlate with higher PoC coverage. Specifically, for CVSS v3.1, PoC coverage was 4.07\% for LOW severity, 9.39\% for MEDIUM severity, 11.27\% for HIGH severity, and 13.94\% for CRITICAL severity. For CVSS v2, PoC coverage was 11.47\% for LOW severity, 19.36\% for MEDIUM severity, and 29.69\% for HIGH severity.}
\textbf{\textit{3)}} Program types. Our measurement of the top-10 open-source and closed-source vulnerability products in CVE shows no clear correlation between PoC coverage and the program types. For instance, {the top 10 open-source products in CVEs have an average coverage rate of 19.72\%, higher than the 10.99\% coverage for closed-source products. However, many open-source products have low PoC coverage (e.g., TensorFlow at 0.50\% and MySQL Server at 1.14\%). Conversely, several closed-source products show high PoC coverage (e.g., iOS at 19.16\% and watchOS at 16.74\%).} 
% due to page limit

\begin{tcolorbox}[size=title,opacityfill=0.1,breakable]
\noindent \textit{\textbf{Remark:}} \textit{{Both the release time and the severity of a vulnerability impact PoC coverage, while whether the vulnerable product is open-source or not has no significant effect. PoC coverage over time follows a distinct trend, consisting of three main phases: the \textit{Plateau Period}, the \textit{Growth Period}, and the \textit{Contrarian Period}. 
% Additionally, there is a clear positive correlation between CVSS threat ratings and PoC coverage.
}}
\end{tcolorbox}

\begin{figure}
  \centering
  \includegraphics[width=0.95\columnwidth]{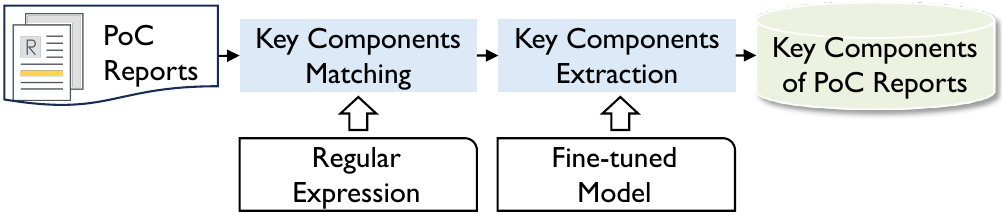}
  \caption{Workflow of \cs{the method of component extraction}.}
  \label{fig:AET.pdf}
\end{figure}

\begin{table}[]
\centering
\caption{Effectiveness evaluation of \cs{the method of key component extraction}. \# Labels refers to the number of labeled key components in the test set.}
\label{tab:AITout}
\scalebox{0.8}{
\begin{tabular}{@{}lcccc@{}}  % Change from 'r' to 'l' for left alignment
\hline
\textbf{Key Component} & \textbf{Precision} & \textbf{Recall} & \textbf{F1-Score} & \textbf{\# Labels} \\ 
\hline
% Title           & 0.77      & 0.65   & 0.71     & 480        \\
% Publish Time      & 0.91      & 0.91   & 0.91     & 1,258       \\
% Author          & 0.80      & 0.60   & 0.69     & 886        \\
% Reference       & 0.94      & 0.99   & 0.96     & 761        \\
% Test Platform    & 0.77      & 0.67   & 0.72     & 471        \\
% Software Version & 0.66      & 0.67   & 0.67     & 2,859       \\
% Trigger Step   & 0.98      & 0.99   & 0.98     & 280        \\
% Verification Oracle   & 0.97      & 0.97   & 0.97     & 88         \\
Title                    & 0.99               & 0.97            & 0.98              & 480                \\
Publish Time             & 0.92               & 1.00            & 0.96              & 1,258              \\
Author                   & 0.92               & 1.00            & 0.96              & 886                \\
Reference                & 0.95               & 0.99            & 0.96              & 761                \\
Test Platform            & 0.91               & 0.97            & 0.94              & 471                \\
Software Version         & 0.87               & 1.00            & 0.93              & 2,859              \\
Trigger Step             & 0.98               & 0.99            & 0.98              & 280                \\
Verification Oracle      & 0.97               & 0.97            & 0.97              & 88                 \\
\hline
\hline
\textbf{Overall}         & 0.91      & 0.99   & 0.95     & 7,083       \\ 
\hline
\end{tabular}
}
\end{table}

%\subsection{Fine-grained Study based on the key components of PoCs}\label{secbert}
% \subsection{Detailed Assessment Focusing on \\key components of PoCs}\label{secbert}
%\subsection{Detailed Assessment on key components of PoCs}\label{secbert}
\subsection{\cs{PoC Report Completeness}}\label{secbert}
%\revise{Update}
% Previous studies have shown that 
%\kx{

%}
% ~\cite{mu2018understanding}. 
Most existing PoC reports are formatted in unstructured plain text, which complicates the task of automatically assessing their \cs{completeness}, especially within a large-scale dataset. 
To enable automated completeness measurement of PoC reports, we first developed a hybrid 
%\underline{k}ey \underline{a}spect \underline{i}dentification \underline{t}ool named \ourtool 
method
to identify and extract the key components within each PoC report, detailed in Section~\ref{secextraKey}. 
After converting PoC reports into structured data, we conducted a fine-grained study to measure their completeness by analyzing the availability of key components.

% \wy{
% \dwj{As it has been proven by the previous research~\cite{mu2018understanding}, the critical information reported greatly affects the success rate of vulnerability recovery}, we define the key components of PoC reports as quantitative indicators for assessing the completeness of PoC reports. }
% \dwj{By defining these key components, it is possible to enable PoC reports to be defined as structured text. Make PoC reports more standardized.}

%\subsubsection{Identification for PoC key components}\label{secextraKey}
\subsubsection{\cs{Key Component Extraction for PoC Report}}\label{secextraKey}
%Based on the key components defined in Section~\ref{sec:keyaspectsdefine}, we developed \ourtool to recognize whether these aspects exist in PoC reports. 
%\ourtool combines the capabilities of Named Entity Recognition (NER) with custom heuristics derived from observations in the raw PoC reports. 
%\dwj{Since it is difficult to assess the quality of \textit{PoC Code} based on its deficiencies, we aim to identify the other 8 key components excluding \textit{PoC Code} from PoC reports.}
% \revise{I think it is difficult to assess the quality of code, which is the main reason. So the current description is incorrect for me}

% \begin{figure}
%   \centering
%   \includegraphics[width=0.99\columnwidth]{img/AITccc.pdf}
%   \caption{Workflow of the \ourtool. }
%   \label{fig:AET.pdf}
% \end{figure}

Figure~\ref{fig:AET.pdf} shows the workflow of \cs{key component extraction}, which takes PoC reports as input and outputs key components \cs{defined in Section~\ref{sec:keyaspectsdefine}}. The tool has 2 main phases, which are \textit{Key Component Matching} and \textit{Key Component Extraction}. 

To extract key components with strict formatting patterns (i.e., \textit{Reference, Trigger Step, and Verification Oracle}), we design regular expressions based on common formatting patterns. 
%observed in the phase of \textit{key component Matching}. 
For example, \textit{Reference} can be matched using URL composition rules. 
Pattern-matching techniques often struggle with identifying components that contain hidden semantics, especially for \textit{Test Platform}, \textit{Software Version}, and \textit{Author}. \cs{Therefore, we take advantage of BERT-NET model to extract these types.}
%the other components of PoC reports. 
%Sun et al.~\cite{sun2023aspect} highlighted the effectiveness of the BERT-NER model~\cite{devlin2018bert} in extracting key components of CVE description. 
%To take advantage of this capability, 
Specifically, we first randomly sampled 2,400 PoC reports from our dataset \cs{, and then
%7 security experts 
3 co-authors annotated}
%to annotate 
these reports using the BIO (Begin, Inside, Outside)~\cite{ramshaw1999text} cross-labeling method. {We divided these PoC reports and their labeled key components into training and test sets in an 8:2 ratio. The training set comprises 26,732 key components from 1,920 PoC reports, while the test set contains 7,083 key components from 480 PoC reports.}  Based on the training set, we fine-tuned the BERT-NER model to identify the remaining components. During fine-tuning, the learning rate is set to 2e-5, the number of epochs to 10, weight decay to 0.01, and the label smoothing factor to 0.1, all of which are intended to enhance the model's generalization capabilities.

As for the effectiveness of our tool, our goal is to determine the presence of each key component in the PoC report rather than extracting the \cs{exact} component. We considered the identification successful if it identified any token within the key component.
\begin{comment}
\dwj{Regrading the evaluation of \ourtool evaluation, we tested \ourtool using five-fold verification~\cite{kohavi1995study}. Our labeled dataset was randomly partitioned into five subsets, and we conducted training and validation five times, each time using a different subset as the test set. The average performance across these five tests was taken as the final result. Our goal was to determine the presence of each key component in the PoC report rather than extracting the entire aspect. Therefore, we considered the identification successful if \ourtool correctly identified any token within the key component. Furthermore, given that a key component may appear multiple times in a PoC report, we calculated \ourtool's performance based on the number of labeled aspects identified.}
% In conclusion, We used all the marked key components as a baseline and considered the identification successful if \ourtool recognized their presence. 
\end{comment}
The experimental result is shown in~\Cref{tab:AITout}, our tool achieved a precision of 0.91, a recall of 0.99, and an F1-score of 0.95. Concretely, components with a common format, such as \textit{Reference} (usually in URL format), or those appearing in a uniform location, such as \textit{Title} (typically at the beginning of the PoC report), performed very well. The \cs{extraction} accuracy of \textit{Title}, \textit{Reference}, \textit{Trigger Step}, and \textit{Verification Oracle} all exceeded 0.95. \cs{Key components} including \textit{Publish Time}, \textit{Author}, \textit{Test Platform}, and \textit{Software Version} are more varied in format and thus relatively challenging to identify and extract. However, our tool still achieved precision above 0.87, recalls above 0.97, and F1-scores above 0.93. We believe that the current performance in extracting key components is sufficient to assess their presence \cs{and help evaluate the completeness of PoC reports}.

\subsubsection{\cs{Assessment of PoC Report Completeness}}\label{secstudykey}
Based on the extracted key components, we further measured the completeness of PoC reports by calculating their presence. Firstly, we analyzed the presence of each key component across a large-scale dataset to determine the overall completeness of the collected PoC reports. Secondly, we compared the presence rates across different databases and examined potential influencing factors. Thirdly, to identify ways to improve the completeness of PoC data, we investigated the possibility of cross-complementation of information in PoC reports for the same vulnerability across multiple datasets.

% \revise{Cannot understand}
\begin{table*}[]
\centering
\caption{Overall percentage of key components extracted from PoC reports. }
\label{tab:overallaspects}
\scalebox{0.7}{
\begin{tabular}{@{}l|c|c|cccc|cc|cc@{}}
\hline
\textbf{} &
  \textbf{\# PoC} &
  \textbf{Presence rate} &
  \multicolumn{4}{c|}{\textbf{Basic Information}} &
  \multicolumn{2}{c|}{\textbf{Environment Deployment Information}} &
  \multicolumn{2}{c}{\textbf{Vulnerability Trigger Information}} \\ \hline
\textbf{\begin{tabular}[c]{@{}l@{}}Database\end{tabular}} &
  \textbf{Total} &
  \textbf{On average} &
  \textbf{Title} &
  \textbf{Author} &
  \textbf{\begin{tabular}[c]{@{}c@{}}Publish  Time\end{tabular}} &
  \textbf{Reference} &
  \textbf{\begin{tabular}[c]{@{}c@{}}Test Platform\end{tabular}} &
  \textbf{\begin{tabular}[c]{@{}c@{}}Software Version\end{tabular}} &
  \textbf{\begin{tabular}[c]{@{}c@{}}Trigger Step\end{tabular}} &
  \textbf{\begin{tabular}[c]{@{}c@{}}Verification Oracle\end{tabular}} \\ \hline
\rowcolor[HTML]{EFEFEF} 
Red Hat Bugzilla        & 176,165 & 80.5\% & 100.0\% & 100.0\% & 100.0\% & 68.0\% & 100.0\% & 100.0\% & 76.2\% & 0.1\%  \\
Talos                   & 1,766   & 79.7\% & 99.5\%  & 99.5\%  & 99.5\%  & 99.8\% & 7.8\%   & 99.8\%  & 76.0\% & 55.7\% \\
\rowcolor[HTML]{EFEFEF} 
Hunter                  & 3,812   & 72.3\% & 100.0\% & 99.2\%  & 100.0\% & 96.4\% & 0.9\%   & 100.0\% & 53.3\% & 28.4\% \\
Exploit Database        & 45,471  & 70.2\% & 100.0\% & 100.0\% & 100.0\% & 70.7\% & 100.0\% & 58.3\%  & 31.8\% & 0.5\%  \\
\rowcolor[HTML]{EFEFEF} 
PHP::Bugs & 80,477  & 69.7\% & 99.9\%  & 100.0\% & 100.0\% & 26.1\% & 89.4\%  & 100.0\% & 42.3\% & 0.0\%  \\
Vulnerability Lab       & 1,516   & 69.5\% & 100.0\% & 100.0\% & 100.0\% & 91.9\% & 1.3\%   & 63.2\%  & 98.2\% & 1.8\%  \\
\rowcolor[HTML]{EFEFEF} 
Chromium       & 10,217  & 69.2\% & 100.0\% & 100.0\% & 100.0\% & 50.2\% & 19.3\%  & 100.0\% & 84.1\% & 0.0\%  \\
CXSecurity              & 40,377  & 66.8\% & 100.0\% & 100.0\% & 100.0\% & 92.3\% & 34.2\%  & 53.9\%  & 53.0\% & 0.6\%  \\
\rowcolor[HTML]{EFEFEF} 
Metasploit              & 2,101   & 62.1\% & 99.9\%  & 99.8\%  & 99.3\%  & 98.0\% & 99.7\%  & 0.1\%   & 0.0\%  & 0.0\%  \\
Packet Storm            & 45,757  & 57.2\% & 100.0\% & 100.0\% & 26.8\%  & 91.4\% & 29.0\%  & 55.9\%  & 53.8\% & 1.0\%  \\
\rowcolor[HTML]{EFEFEF} 
Seebug                  & 41,565  & 53.7\% & 100.0\% & 22.0\%  & 100.0\% & 96.2\% & 11.0\%  & 55.5\%  & 44.0\% & 0.9\%  \\
Openwall                & 6,970   & 22.7\% & 3.3\%   & 3.9\%   & 3.1\%   & 83.5\% & 5.1\%   & 26.2\%  & 55.9\% & 0.4\%  \\ \hline
% \hline
\rowcolor[HTML]{EFEFEF} 
\textbf{Overall} &
  \textbf{456,194} &
  \textbf{70.3\%} &
  \textbf{98.5\%} &
  \textbf{91.4\%} &
  \textbf{91.2\%} &
  \textbf{68.4\%} &
  \textbf{72.3\%} &
  \textbf{81.6\%} &
  \textbf{58.0\%} &
  \textbf{0.8\%} \\ \hline
\end{tabular}
}
\end{table*}

\noindent\textbf{Presence of key components on the large-scale dataset.}
% Since \textit{PoC Code} represents a specific implementation of a PoC and is primarily dependent on the implementation method rather than the completeness of the PoC reports, we focus on the other 8 key components to evaluate the completeness of PoC reports. 
The last row of~\Cref{tab:overallaspects} shows that the average presence rate is 70.3\%. Note that a subset of 14,787 PoC reports, characterized by containing multiple files within complex catalog structures and scattered text, was excluded from our analysis. This subset represents a small fraction (3.14\%) of the total dataset and thereby does not substantially impact the overall results. 
After this exclusion, the analysis proceeded with the remaining 456,134 PoC reports. 
The \cs{components of the \textit{Basic Information} category} are relatively intact, with \textit{Title}, \textit{Author}, and \textit{Publish Time} all having a presence rate of over 90\%, and \textit{Reference} present at almost 70\%. 
However, the components of the \textit{Environment Deployment Information} and \textit{Vulnerability Trigger Information} categories have an overall low probability of existence. 
The \textit{Test Platform} and \textit{Software Version} are only present in 72.3\% and 81.6\%. 
\textit{Trigger Step} contains the specific steps needed during reproduction and should also be detailed in the PoC report. However, it is present in only 58.0\% of the cases. 
The \cs{component} with the lowest presence is \textit{Verification Oracle}, a crucial indicator of whether a vulnerability has been successfully triggered, appearing in only 0.8\% of the overall dataset. 
% \dwj{We observed that no existing platforms explicitly require the submission of a \textit{Verification Oracle} as part of the process. Furthermore, \textit{Verification Oracle} can take various forms, including textual descriptions, images, and videos; however, some platforms do not support the uploading of images and videos. These limitations likely contribute to the low prevalence of this critical component. }
Our analysis suggests a twofold reason. 
Firstly, no existing \cs{databases} explicitly require the inclusion of a \textit{Verification Oracle} as part of the submission, potentially leading to its infrequent use. 
Secondly, while a \textit{Verification Oracle} can be presented in various formats such as images or videos, many \cs{databases} typically limit uploads to text-only. This restriction hampers the ability of PoC authors to provide a comprehensive demonstration of vulnerability, further reducing the presence of \textit{Verification Oracle}.

\noindent\textbf{Presence of key components across different databases.} 
\Cref{tab:overallaspects} reflects that the presence of key components varies significantly among databases. 
The databases with the highest key component presence rates are Bugzilla (80.5\%), Talos (79.7\%), and Hunter (72.3\%). These \cs{databases} commonly feature pre-defined templates for submitting PoC reports. For instance, Bugzilla requires submitters to fill in 29 predefined fields. Specifically, the template defined by Bugzilla contains two fields ``OS'' and ``Version'', which can be mapped into our defined key components \textit{Test Platform} and \textit{Software Version}. Consequently, the presence rates of these two \cs{components} in Bugzilla is 100\%, significantly higher than those in other databases. In contrast, the \cs{database} with the lowest average presence of key components is Openwall (22.7\%), which allows submissions in any format. Generally, well-defined submission templates may enhance the presence of key components in PoC reports. 

{We also measured how the presence of key components varies across PoC languages, program types, and CWEs. The results indicate that these factors did not have a significant impact.}

\noindent{\textbf{Possibility of information fusion from multiple databases.}} 
\Cref{sec: poc-perspective} has elaborated that there are many instances of multiple PoC reports pointing to the same CVE, primarily occurring across different databases. 
To further enhance the data completeness of the current PoC report, we investigate the potential of fusing information from multiple databases. 

To this end, we explored whether the key components in the PoC reports of Exploit DB, CXSecurity, and Seebug can complement one another if multiple PoC reports point to the same CVE ID, for the fact that the three databases have the highest number of PoC reports pointing to the same CVEs among them. 
As shown in~\Cref{tab: multisource}, the presence of key components has improved considerably from 51.8\% to 73.6\% after performing data complements between the three databases. This suggests that 
%the cross-completion of 
\cs{the complemented}
PoC reports \cs{based on} multiple sources can significantly improve the \cs{completeness} of existing PoC reports on large datasets. 
% From this result, it can be seen that we can use multi-source information fusion methods for PoC data complementation, thus improving the quality of existing PoC data.

% \begin{table*}[]
% \caption{Results of multi-source information complementation. }
% \label{tab: multisource}
% \resizebox{1\linewidth}{!}{
% \begin{tabular}{l|ccccccccl}
% \hline
%                          & Title    & Author   & Time     & Reference & Test Platform & Software Version & Trigger Method & Verification Oracle & Average \\ \hline
% Origin presence          & 100.0\%  & 74.8\%  & 100.0\% & 95.3\%   & 45.0\%      & 74.9\%         & 42.9\%       & 1.0\%        & 63.6\% \\
% After refusion & 100.0\% & 100.0\% & 100.0\% & 100.0\%  & 99.8\%      & 100.0\%         & 66.5\%       & 2.0\%        & 83.5\% \\ \hline
% \end{tabular}
% }
% \end{table*}

\begin{table}[]
\centering
\caption{Results of multi-source PoC information fusion \cs{using} three databases \cs{(Exploit DB, CXSecurity, and Seebug)}.}
\label{tab: multisource}
\scalebox{0.75}{
\begin{tabular}{l|ccccc|c}
\hline
\textbf{} &
  {\textbf{Reference}} &
  {\textbf{\begin{tabular}[c]{@{}c@{}}Test\\  Platform\end{tabular}}} &
  {\textbf{\begin{tabular}[c]{@{}c@{}}Software\\Version\end{tabular}}} &
  {\textbf{\begin{tabular}[c]{@{}c@{}}Trigger\\  Step\end{tabular}}} &
  {\textbf{\begin{tabular}[c]{@{}c@{}}Verification\\Oracle\end{tabular}}} &
  \textbf{Average} \\ \hline
Origin &
  95.3\% &
  45.0\% &
  74.9\% &
  42.9\% &
  1.0\% &
  51.8\% \\
Fusion &
  100.0\% &
  99.8\% &
  100.0\% &
  66.5\% &
  2.0\% &
  73.6\% \\ \hline
\end{tabular}
}
\end{table}

\begin{tcolorbox}[size=title,opacityfill=0.1,breakable]
\noindent \textit{\textbf{Remark:}} 
\textit{{The overall presence of key components of PoC reports is 70.3\%.
% , indicating that many PoC reports are missing important information needed to reproduce.
Components related to {Basic Information} are relatively complete, while others are less present. 
{Meanwhile, PoCs pointing to the same CVE across databases can complement one another with the necessary information, improving the completeness of existing PoC reports.}
}}
\end{tcolorbox}

% \subsection{Reproduction Study Based on PoC Reports}\label{secrepro}

\subsection{\cs{PoC Reproducibility}}\label{secrepro}
As discussed in~\Cref{secbert}, there is still considerable room for improvement in the completeness of the PoC reports, 
we thereby explore how far the existing PoC reports are from being able to reproduce a vulnerability. 
% Although mu et al.~\cite{mu2018understanding} also conducted research related to vulnerability recurrence, we clarify that our work is quite different from theirs. The main differences are: \ding{182} Different research objects. Their study is on vulnerability reports, which may not contain PoCs, while our study is on PoC reports. \ding{183} Broader sampling. They only sampled vulnerabilities in the Linux kernel, while we sampled randomly from the large PoC dataset we collected in the hope of providing insights into a broader dataset. \ding{184} Deeper insights. They explored what key components are more meaningful for vulnerability reproduction, focusing on the importance of completing vulnerability reports and staying at the level of calling for high-quality data. We, on the other hand, focus on how to help improve the reproduction success rate in real-world applications based on existing PoC reports, focusing more on the technical aspects. 

% \subsubsection{Reproduction Study Overview}
% \subsubsection{Overview of Reproduction Study}

\subsubsection{Assessment of PoC Reproducibility}

\begin{figure*}
	\centering
	\subfloat[Workflow of \cs{assessment of PoC reproducibility}.]{\includegraphics[width=1.3\columnwidth]{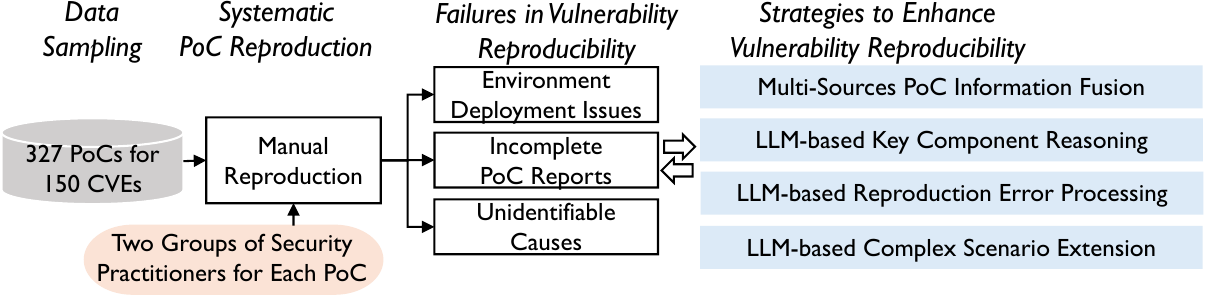}\label{fig: reprooverview.pdf}}
        \hspace{3pt}
	\subfloat[\cs{Reproduction} results \cs{following the enhancements made by the participants.}]
 {\includegraphics[width=.66\columnwidth]{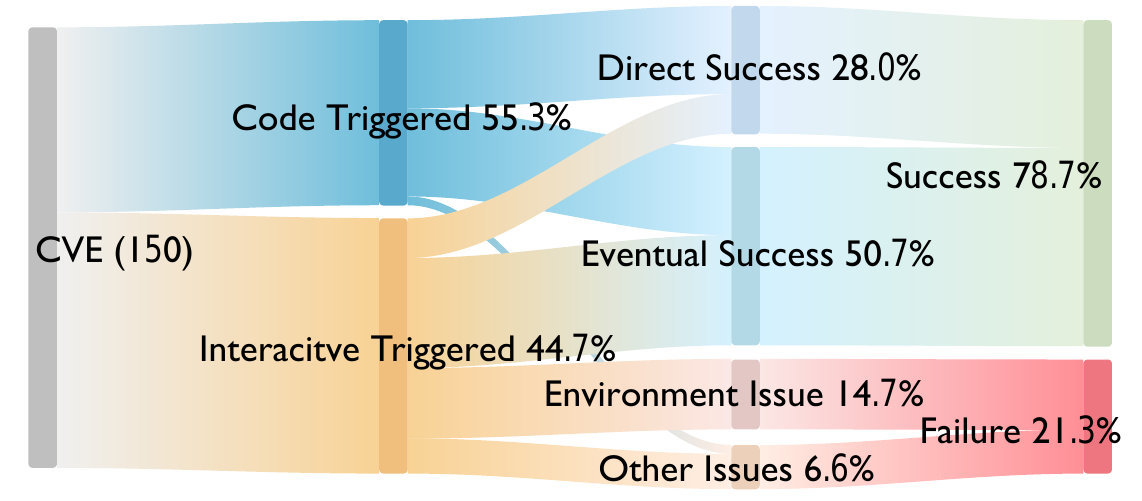}\label{fig:sankey.pdf}}
	\caption{Workflow of \cs{assessment of PoC reproducibility and its result}.}
\end{figure*}

% Figure~\ref{fig: reprooverview.pdf} shows the overview of the reproduction study. First, we sampled 100 PoC reports as our reproduction dataset, including 50 logic vulnerabilities~\cite{felmetsger2010toward} and 50 memory-corrupting vulnerabilities~\cite{szekeres2013sok}. These two vulnerability types could almost cover all common vulnerabilities. Next, After deploying the environment as described in the PoC report, we then invited seven long-time practitioners in the security space to manually reproduce these PoC reports. We found that the probability that a vulnerability can be directly reproduced based on an existing PoC report is pretty low, so we propose three main methods to assist PoC reproduction. The three methods are tuning PoC code, multi-sources fusion of PoC information, and using LLM prior knowledge assistance. 
\cs{As shown in \Cref{fig: reprooverview.pdf}, to assess the effectiveness of the PoC reports, we conducted a human study by reproducing the vulnerability based on the corresponding PoC reports.}
%\Cref{fig: reprooverview.pdf} shows the workflow of \cs{our study.}
% First, we sampled 100 CVE IDs and searched for their PoC reports as a reproduction dataset. \cmt{don't know whether all 100 cves have poc here.}
% Next, after deploying the environment described in the PoC reports, we invited seven longtime security professionals to manually reproduce these PoC reports. 
% We found that the probability of directly reproducing vulnerabilities based on existing PoC reports is quite low, so we proposed three main approaches to assist PoC reproduction. The three methods are code debugging, fusing PoC information from multiple sources, and using LLM a prior knowledge assistance.
%Concretely, we sampled 100 CVEs and the corresponding PoC reports from 49,667 CVEs {that can be queried for the corresponding PoC report} as a reproduction dataset. 
%We invited 8 longtime security experts to manually reproduce these PoC reports. 

% Since it is challenging to directly reproduce vulnerabilities by relying on existing PoC reports, we proposed three main approaches to assist PoC reproduction. 
% The three methods are code debugging, fusing PoC information from multiple sources, and using LLM a prior knowledge assistance.

%\smallskip
\noindent\textbf{\cs{Dataset used in human study.}}
% We aimed to ensure representativeness in our sample of vulnerabilities.  
We randomly sampled \dwj{150} CVE vulnerabilities, each accompanied by the available corresponding PoC reports. 
Among these, \dwj{87} CVEs are associated with a single PoC report, while the remaining \dwj{63} CVEs have multiple PoC reports, resulting in a dataset of 150 CVEs and \dwj{327} PoC reports. 
The CVEs span from 2003 to 2024, with PoCs sourced from 10 different databases. 
%\dwj{A total of 23 CVE entries describe vulnerability types, covering 18 distinct types of vulnerabilities.} %Additionally, 48 CVE entries describe vulnerable products, encompassing 41 unique products.
\cs{The CVEs we sampled include \dwj{32} vulnerability types in total, covering all common types such as race condition, buffer overflow, and privilege escalation. The sufficient variety of types ensures the generalizability of our experimental conclusions.}

\noindent\textbf{Participant recruitment.}
\cs{We recruited ten participants, including 2 postdoctoral researchers, 4 PhD students, and 4 Master's students. They all have over 3 years of experience in the field of security and have been actively involved in research related to vulnerabilities and PoCs.}

% \smallskip
% \noindent\textbf{Systematic expert reproduction.} 

\noindent\textbf{Experiment procedures.}
To minimize manual bias, we divided the ten participants into two groups, each tasked with reproducing all \dwj{150} vulnerabilities. Each participant reproduces \dwj{30} vulnerabilities, and the reproduction results of the two groups are compared for consistency. In cases of inconsistency, a discussion is held, followed by another attempt at reproduction, until consistent results are achieved: we employ a two-step process: \textbf{\textit{1)}} Participants are initially required to use only the information in the PoC reports, thereby directly measuring their reproducibility. 
\textbf{\textit{2)}} Next, for vulnerabilities unable to reproduce directly, participants use their domain knowledge and \cs{available} tools to aid the reproduction. 

\noindent{\textbf{\cs{Experiment results.}}}
\cs{Only 42 CVE vulnerabilities were directly reproduced successfully based on the initial PoC reports. The reasons for initial failure include 47 vulnerabilities with incomplete PoC reports, 16 with environment deployment issues, 4 with PoC execution errors, 2 with unexpected outputs, and 2 with unidentified causes.}
\cs{Among the remaining \dwj{108} CVE vulnerabilities, \dwj{76} were successfully reproduced following enhancements made by the participants, including data augmentation and code debugging, as detailed in Section~\ref{sec:enhancerepro}.}
\cs{Figure~\ref{fig:sankey.pdf} shows that \dwj{78.7\%} \dwj{(118/150)} of CVEs were successfully reproduced through the PoC reports we collected, leaving the other \dwj{32} vulnerabilities failing to be reproduced eventually.}

% Interestingly, some of the direct success cases were facilitated by PoC reports that provided complete reproduction environments. For instance, the PoC report for CVE-2021-44228 (Log4Shell)~\cite{GitHubcy6:online} includes a Docker file with an exploit demonstration on a Tomcat 8 server. Practices like this prove immensely helpful for users, especially in cases where environment deployment is inherently troublesome or difficult.

% We first discuss the relationship between reproducibility and potential factors in this section. 

% \begin{figure*}[!t]
%     \begin{minipage}{0.58\linewidth}
%         \centering
%         \captionof{table}{Reproduction result related to trigger forms.}
%         \resizebox{1\linewidth}{!}{
%         \begin{tabular}{ccccc}
%             \toprule
%             \textbf{\makecell{Trigger Form}} & \textbf{\makecell{Direct\\ Success}} & \textbf{\makecell{Eventual\\ Success}} & \textbf{Failure} & \textbf{\makecell{Avg. Efforts\\ (man-hours)}} \\ \midrule
%             \textbf{Code-triggered} & 20 (47.62\%) & 20 (47.62\%) & 2 (4.76\%) & 3 \\
%             \textbf{Interactive-triggered} & 9 (15.52\%) & 25 (43.10\%) & 24 (41.38\%) & 5.5 \\ \bottomrule
%         \end{tabular}}
%         \label{tab: triggerresult}
%     \end{minipage}
%     \hspace*{0pt}
%     \centering
%     \begin{minipage}{0.35\textwidth}
%         \centering
%         \includegraphics[width=\linewidth]{img/6-2-datasource.pdf}
%         \captionof{figure}{Reproducibility vs. \cs{databases}.}
%         \label{fig: 6-2-datasource.pdf}
%     \end{minipage}
% \end{figure*}

\begin{table}
        \centering
        \captionof{table}{Reproduction result related to trigger forms.}
        \scalebox{0.75}{
        \begin{tabular}{ccccc}
            \hline
            \textbf{\makecell{Trigger Form}} & \textbf{\makecell{Direct\\ Success}} & \textbf{\makecell{Eventual\\ Success}} & \textbf{Failure} & \textbf{\makecell{Avg. Efforts\\ (man-hours)}} \\ \midrule
            \textbf{Code-triggered} & 34 (51.0\%) & 45 (54.2\%) & 4 (4.8\%) & 3 \\
            \textbf{Interactive-triggered} & 8 (11.9\%) & 31 (46.3\%) & 28 (41.8\%) & 5.5 \\ \hline
        \end{tabular}}
        \label{tab: triggerresult}
\end{table}
% \begin{figure}[]
%   \centering
%   \includegraphics[width=0.8\linewidth]{img/6-2-datasource.pdf}
%   \caption{Reproducibility vs. databases.\revise{update}}
%   \label{fig: 6-2-datasource.pdf}
% \end{figure}

% \begin{figure}[]
%   \centering
%   \includegraphics[width=0.99\linewidth]{img/Overall reproduction result.pdf}
%   \caption{Overall reproduction result.}
%   \label{fig:sankey.pdf}
% \end{figure}

% \begin{table}[]
% \caption{Reproduction result related to trigger forms. }
% \label{tab: triggerresult}
% \resizebox{1\linewidth}{!}{
% \begin{tabular}{ccccc}
% \toprule
% \textbf{\makecell{Trigger Form}  }        & \textbf{\makecell{Direct\\ Success}} & \textbf{\makecell{Eventual\\ Success}} & \textbf{Failure}     & \textbf{\makecell{Avg. Efforts\\ (man-hours)}} \\ \midrule
% \textbf{Code-triggered }       & 20 (47.62\%)   & 20 (47.62\%)     & 2 (4.76\%)  & 3             \\
% \textbf{Interactive-triggered} & 9 (15.51\%)    & 25 (43.10\%)     & 8 (13.79\%) & 5.5           \\ \bottomrule
% \end{tabular}
% }
% \end{table}

% \smallskip
%\noindent\textbf{Reproducibility vs. Trigger forms.} 
% As discussed in Figure~\ref{fig:form} before, there are two trigger forms for vulnerability reproduction, i.e., \textit{code triggered} and \textit{interactive triggered}. 
%We observed a correlation between vulnerability reproducibility and the trigger forms of PoC reports. 

We aim to observe the impact of different trigger types on reproducibility.
As shown in~\Cref{tab: triggerresult}, the success rate for \textit{code triggered} reproductions is higher than that for \textit{interactive triggered}. Specifically, the direct success rate for \textit{code-triggered} approaches is \dwj{51.0\%}, significantly higher than the \dwj{11.9\%} for \textit{interactive-triggered} approaches.  
Furthermore, our findings suggest that reproductions initiated by \textit{code triggered} not only achieve a higher success rate but also require less time. Excluding the time spent on setting up the environment, \textit{interactive triggered} requires an average of 5.5 man-hours per vulnerability, 
%significantly 
more than the 3 man-hours required on average for \textit{code triggered} reproductions. 
% In addition, we recorded the manual time required to reproduce each CVE entry, and after excluding the time consumed by the environment deployment process, we found that the average time consumed by manually triggering a vulnerability is 2.5 man-hours, which is much higher than the average time required by code triggering a vulnerability (1 man-hour). 
This efficiency may be attributed to the fact that PoC reports relying on \textit{interactive triggering} often assume that potential readers have a certain level of expertise and familiarity with the targeted vulnerable system, thereby omitting critical details needed to reproduce the CVEs. 
For example, the PoC for CVE-2012-5346 only provides the payload without the necessary attack path, demanding extensive extra analysis.

% \smallskip
%\noindent\textbf{Reproducibility vs. databases.} 
% As discussed in~\Cref{secbert}, the completeness of PoC reports is related to their databases. 
% Hence, we further observe the reproducibility of PoC reports across various sources. 
% \dwj{As shown in~\Cref{fig: 6-2-datasource.pdf}, among the sampled PoCs, Databases that are more complete in key aspects have a relatively higher reproduction success rate. Specifically, the reproduction rate for Hunter was as high as 100\%, while Talos had the highest direct reproduction success rate of 87.5\%.} 

\begin{tcolorbox}[size=title,opacityfill=0.1,breakable]
\noindent \textit{\textbf{Remarks:}}
\textit{
% \dwj{Only a small portion of vulnerabilities (29\%) were successfully reproduced directly using the initial PoC reports. A larger proportion (45\%) required further enhancements by participants before they could be successfully reproduced. Ultimately, 26\% of the vulnerabilities could not be reproduced.}
\cs{Only \dwj{28.0\%} of the vulnerabilities were directly reproducible using initial PoC reports. A larger part, \dwj{50.7\%}, required additional enhancements from the participants to achieve successful reproduction. In the end, \dwj{21.3\%} of the vulnerabilities remained unreproducible.}
}
% {\textit{{\dwj{The reproducibility of vulnerability recurrence is somewhat related to the form in which the PoC report is triggered and the source of the data. 
% Our observation is: \ding{182} The success rate for code-triggered reproductions
% is higher than that for interactive triggered. \ding{183} Vulnerabilities with high-quality PoC reports have a higher success rate of reproduction.}
% }}}
\end{tcolorbox}

\subsubsection{Strategies to Enhance Reproducibility}\label{sec:enhancerepro}
% Based on the results of the reproduction and the expert experience of the reproduction process, 
% we summarized the cases that were not directly successfully reproduced, but finally succeeded through continuous attempts, and concluded three efficient and commonly used suggestions to assist vulnerability verification and improve the success rate of vulnerability verification.
As illustrated in~\Cref{fig:sankey.pdf}, many vulnerability PoCs cannot be reproduced successfully on the first attempt but are eventually reproduced after further processing. We identified three key strategies that significantly enhance the reproducibility as follows: 
% We differentiate between cases where CVE entries were reproduced directly and those that required additional expertise or techniques for successful validation.
% As shown in~\Cref{fig:sankey.pdf}, many vulnerability PoCs cannot be reproduced successfully directly, but are eventually reproduced after further processing. 
% Based on the outcomes of the vulnerability reproduction efforts, we identified key strategies that significantly improve the success rate of reproducing vulnerabilities. We distinguish between cases where CVE entries were reproduced directly and those that required additional expertise or techniques for successful validation. 
% We offer three practical recommendations to aid in effective vulnerability reproduction and verification:

%\smallskip
\noindent\textbf{Multi-sources PoC information fusion.}
Given the noted relatively low completeness of existing PoC reports (Section~\ref{secbert}) and the presence of multiple PoC reports for the same CVE 
% (Section~\ref{secoverlap})
, we identified a significant opportunity to improve the \cs{completeness} of vulnerability PoCs. 
% By infusing the key components across various PoC reports associated with the same vulnerability, we could obtain a more unified and detailed depiction of the vulnerability reproduction. 
{For example, the two PoC reports for CVE-2014-0160 are incomplete, }lacking critical information necessary for reproduction such as \textit{Test Platform}, \textit{Trigger Step}, and \textit{Verification Oracle}. 
As shown in Figure~\ref{fig: figmulti.pdf}, after key components fusion based on another PoC report retrieved from our large-scale dataset, we obtained a PoC report that is complete in key components and more reproducible.
More importantly, our extensive PoC dataset includes an additional 38 PoC reports for this CVE, each providing valuable details that the CVE description does not cover. 
% With our aspect identification tool \ourtool and our dataset, we can automate data enhancement for PoCs, improving the feasibility of reproducing and verifying vulnerabilities. 
By applying this strategy, we enhanced the information for the PoCs of 27 CVEs and successfully reproduced them. 

\begin{figure}[]
  \centering
  \includegraphics[width=0.45\textwidth]{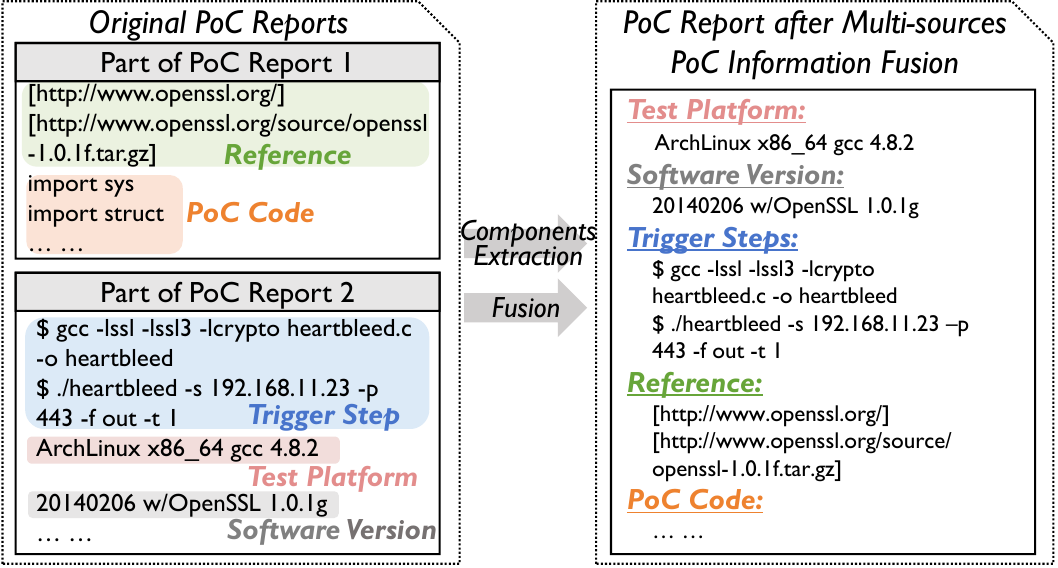}
  \caption{Example of multi-source PoC information fusion.}
  \label{fig: figmulti.pdf}
\end{figure}

%\smallskip
\noindent\dwj{\textbf{LLM-assisted key component reasoning.}
LLMs have demonstrated exceptional capabilities when aiding in security tasks like vulnerability identification, verification, and remediation. 
We also observed that LLMs are particularly beneficial for vulnerability reproduction in completing the missing components of PoC reports. 
We tested the reasoning capabilities of state-of-the-art LLMs using Chain-of-Thought reasoning~\cite{wei2022chain} and selected the best-performing LLMs for key components reasoning. 
For instance, in the case of CVE-2019-0221 affecting Apache Tomcat, the existing PoC report lacked details about the \textit{Test Platform}. 
ChatGPT-4o-latest recommends using Ubuntu 20.04 to reproduce this vulnerability, based on the \textit{Software Version} (Apache Tomcat 9.0.0, which supports installation on Ubuntu 20.04) and the specific write-up in the \textit{Trigger Step} (the use of a Shell script). 
By using this strategy, we successfully reproduced 37 CVE entries that initially failed. 
% We provide the specific prompt for ChatGPT-4o-latest on~\cite{reproductionresult}.}
}

\noindent\textbf{LLM-based reproduction error diagnosis.} 
Various errors may be encountered during the reproduction process, such as code errors and unexpected output. 
It usually relies heavily on manual inspecting to analyze the root cause and resolve the error, which is a labor-intensive and time-consuming process. 
Another observation is that LLMs can efficiently assist in deducing the root cause of the error and suggesting adjustments based on the reported error information. For instance, 
in our attempts to reproduce the CVE-2016-5195 ``Dirty Cow'' vulnerability~\cite{CVERecor29:online}, we initially faced error messages while processing a system file using a C script. After feeding the error message to ChatGPT-4o-latest, it quickly pinpointed a compiler version incompatibility issue among the many possible potential problems that were not mentioned in the PoC report, which was not initially considered by the participants. 
Additionally, Linux kernel-related vulnerabilities generally require debugging trials and take a long time to reproduce, but reproducing CVE-2016-5195 took only 1.5 man-hours with the assistance of LLMs. 
By applying this strategy, we successfully reproduced 9 vulnerabilities that did not work at the beginning.

\noindent\textbf{LLM-assisted PoC scenario adaptation.}
Some vulnerability reproduction failures stem from PoCs designed for specific, often obsolete, deployment scenarios. While the core vulnerability may persist, the original exploit’s dependencies or environmental assumptions (e.g., framework versions, libraries, or OS configurations) can render it non-executable in modern or alternative contexts.
By analyzing the original PoC’s intent (e.g., payload delivery), LLMs can rewrite the exploit logic into a functionally equivalent but environment-deployable form (eliminating outdated dependencies, migrating frameworks, or adjusting API calls). 
The Metasploit-based PoC for CVE-2017-7494 fails on modern systems due to dependency conflicts. Using ChatGPT-4o, we translated it into a standalone Python script that replicates the attack logic (e.g., SMB payload injection) without Metasploit, resolving version incompatibilities. This adaptation enabled successful reproduction, whereas the original PoC failed. 
Applied to three PoCs previously unreproducible, this strategy demonstrated consistent success in bypassing environment-specific barriers.
% , confirming its viability for improving exploit reproducibility.

% For instance, the original PoC for CVE-2017-7494 was implemented using the Metasploit framework. Due to dependency and version incompatibilities, this PoC is no longer executable in its original form. Using ChatGPT-4o-latest, we successfully translated the PoC into a standalone Python script, effectively decoupling it from the Metasploit framework and eliminating environment-related deployment issues.
% Applying this strategy, we have successfully rewritten and reproduced three previously non-functional PoCs, demonstrating the feasibility and effectiveness of LLM-assisted PoC transformation.}

\begin{tcolorbox}[size=title,opacityfill=0.1,breakable]
\noindent \textit{\textbf{Remarks:}}
% Based on our research on vulnerability reproduction, 
% \dwj{Various technical approaches, including data fusion, LLM assistance for completing key components, and LLM-based reproduction error diagnosis, could significantly improve reproduction success rates.}
{\textit{{Various technical approaches and existing tools, including data fusion, LLM assistance for completing key components, LLM-based reproduction error diagnosis, and LLM-based reproduction error diagnosis could improve reproduction success rates.
}}}
\end{tcolorbox}

\subsubsection{Failures in Vulnerability Reproduction}\label{secreprofail}
% Having previously discussed the relationship between success rates and individual factors, we still want to recount the reasons for our trigger failures. 
As shown in~\Cref{img/6-2-failure.pdf}, 21.3\% (32/150) of sampled vulnerabilities failed to be reproduced after adopting enhanced strategies. These eventual failures can be attributed to 3 primary causes: 
\begin{itemize}[leftmargin=5pt] 
    \item Environment deployment issues (22 cases): 
    The majority of reproduction failures were due to the inability to deploy the environment on which vulnerability reproduction depends. 
    6 PoCs involved closed-source software and therefore could not deploy the environment. 
    While 8 PoCs were for open-source software, the vulnerable version was no longer available. 
    The final 2 PoCs involve the environment setup requiring 2 deprecated dependency packages that are now unavailable. 
    
    \item Incomplete reports (8 cases): 
    These reproduction failures arise because the reports lack the necessary information. 
    Concretely, 7 PoCs lack a specific \textit{Trigger Step}, which was particularly problematic for closed-source software.
    % , where our efforts to develop a complete \textit{Trigger Step} using the limited information available were unsuccessful
    PoC for CVE-2003-1506 lacks a specific \textit{Software Version} and is dated, making the vulnerability unable to be reproduced.
    % This vulnerability is an SQL injection previously present on the Consornet~\cite{Autonomo71:online} website. \cmt{why miss software version??? you now state is that the vuln has been fixed! vuln version has also been disappeared.}
    % However, it has been fixed and current server configurations within the Consornet website effectively detect and block anomalous SQL injection attempts, as evidenced by a 403 error response during our reproduction process. 
    % This indicates successful remediation of the reported issue. 
    \item Unidentifiable causes (2 cases): 2 reproductions (CVE-2003-1506 and CVE-2016-4484) failed for reasons that could not be determined. Our two groups of \cs{participants} tried all possible methods but still failed to reproduce. 
\end{itemize}

\begin{figure}
    \centering
    \includegraphics[width=0.8\columnwidth]{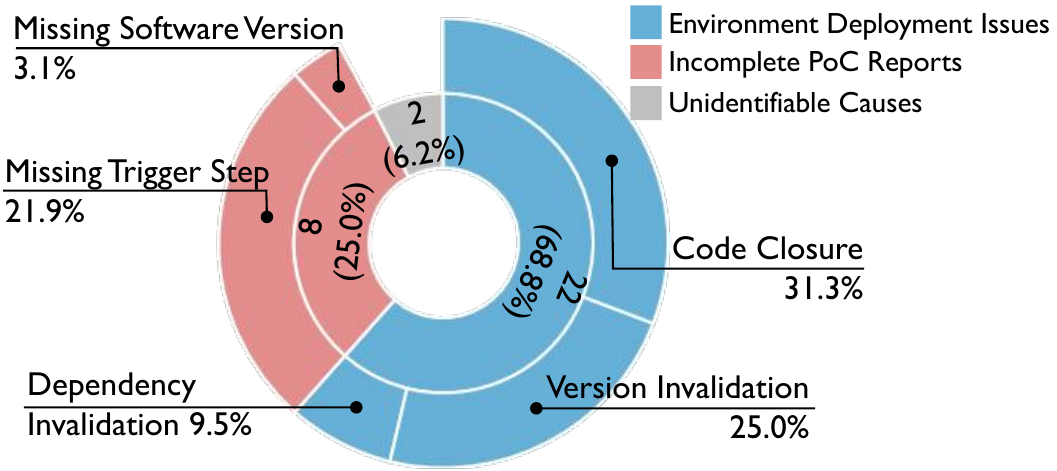}
      \caption{Reasons for failure to reproduce a vulnerability.}
      \label{img/6-2-failure.pdf}
\end{figure}
% \begin{tcolorbox}[size=title,opacityfill=0.1,breakable]
% \noindent \textit{\textbf{Remarks:}} 
% {\textit{{
% \dwj{Based on our research on vulnerability reproduction, we found that factors mainly including the completion of key components significantly affect the success rate. Various technical approaches including LLM assistance and data fusion could significantly improve reproduction success rates. Additionally, issues with environment deployment are also important factors affecting the success rate.}
% }}}
% \end{tcolorbox}

\begin{tcolorbox}[size=title,opacityfill=0.1,breakable]
\noindent \textit{\textbf{Remarks:}} {\textit{{
% Based on our research on vulnerability reproduction, 
{The eventual reproduction failures are mainly due to environment deployment issues, such as code closure, version invalidation, dependency invalidation.}
}}}
\end{tcolorbox}

% \revise{Remarks for 4.4}
% \subsection{Obse  rvation and lessons}

% \textbf{The importance of \textit{Record}}

\section{Application and Implication}
\subsection{Application}\label{sec:data-application}
% To facilitate the research and application of PoC data, we have organized and standardized the format of the 470,921 PoC reports collected.
% %, and have made them open source on our website \url{www.sctruster.com/cybersploit/}. 
% Furthermore, we have also implemented an component-level PoC search strategy, making it easier to find target PoCs. %The primary design and search features of our website can be found in~\Cref{sec:website-intro}.
\subsubsection{PoC Retrieval}
It is essential to retrieve PoC reports based on existing information since vulnerability and PoC information are scattered across multiple databases.
We begin by introducing the limitations of current PoC data platforms as follows: 
\textbf{\textit{1)}} As discussed in~\Cref{sec:collectpoc}, existing PoC reports are scattered across multiple databases, which poses challenges for PoC retrieval. 
\textbf{\textit{2)}} Most existing PoC databases support retrieval only by \textit{CVE ID} or \textit{Title}, which may not align well with practical user needs. Vulnerabilities are not archived uniformly, and \textit{CVE ID} is just one of the more commonly used indices. Furthermore, some PoC reports lack the correct link to the corresponding vulnerabilities, as discussed in~\Cref{secPoCCoverage}. Regarding \textit{Title}-based retrieval, \textit{Title} contains only a small portion of information, omitting other critical details, which makes it difficult to accurately match the target PoC. 

To bridge these gaps and facilitate the research and application of PoC data,
%we have organized and standardized the format of the 470,921 PoC reports. Furthermore, 
we have implemented a component-level PoC search strategy, facilitating efficient and convenient retrieval of PoC reports.
Users can retrieve using various keywords including \textit{Author}, \textit{Software Version}, \textit{Test Platform} which we have defined as key components in~\Cref{sec:keyaspectsdefine}. This versatility in search parameters significantly aids users in finding the specific PoCs they require. 
% The component-level search strategy also contributes to downstream tasks, particularly in PoC augmentation and PoC linking. 
% For the PoC augmentation task, when a single PoC report does not contain sufficient information to reproduce a vulnerability, the missing key components can be supplemented by other related PoC reports, such as those pointing to the same CVE ID. Moreover, the multi-key component level search approach provides more similar PoC reports, which helps to enhance the PoC information and improve the success rate of vulnerability reproduction. 
% For the PoC linking task, we can link vulnerabilities and PoC reports by assessing the degree of overlap of the key components when the PoC report omits its corresponding vulnerability information. 
% {The primary design and search features of our website are detailed on~\cite{dataapplication}. We are also in the process of building a multi-faced search for our website based on the already implemented aspect-level search capabilities. }

\subsubsection{Key Component Extraction}
As discussed in~\Cref{secbert}, most of the current PoC reports are semi-structured and unstructured texts. However, as reflected in~\cite{mu2018understanding}, there is a critical need for automated mechanisms to extract and collect useful information from PoC reports. The key component extraction tool aids in alleviating this problem and can automatically parse PoC reports with structured key components.
Furthermore, a number of future research studies can be carried out based on the key components extracted by our tool. For example, users can automate the fusion of complementary PoC reports from multiple sources, and users could develop automated PoC verification environment deployments and semi-automated PoC verification methods with the help of the extracted key components.

\subsection{\cs{Implications}}
\subsubsection{For PoC Developers}
%Based on our findings, we propose the following recommendations:
%We propose the following recommendations.
We recommend:
\textbf{\textit{1)}} \textit{Explicitly state the target vulnerability in the PoC report.} As discussed in~\Cref{secstudykey}, many PoC reports lack links to their corresponding vulnerabilities, hindering their practical application. PoC developers should explicitly state the target vulnerability when submitting a PoC report, especially including vulnerability identifiers such as CVE IDs.
\textbf{\textit{2)}} \textit{Improve the completeness of PoC reports by providing key components.}
As depicted in~\Cref{secstudykey}, existing PoC reports generally lack 30\% key components that are critical for reproduction, especially for environment setup and trigger-related information. Developers are urged to provide comprehensive details since it will significantly improve the practical usability of PoC reports. 
% avoid subjective descriptions to enhance the practical usability of PoC reports. 
\textbf{\textit{3)}} \textit{Facilitate reproduction with ready-to-use environments.} Our study in~\Cref{secreprofail} shows that many failures in reproducing vulnerabilities arise from issues in environment setup. 
Notably, among 29 cases of direct success in~\Cref{secrepro}, we observed that several PoC reports provided pre-configured specifically tailored environments (e.g., docker containers). 
Hence, PoC developers are encouraged to consider this practice where feasible to enhance the reliability of reproducing vulnerabilities. 
\dwj{\textbf{\textit{3)}} \textit{Be more responsible for PoC reports that are publicly available.} Many PoC reports are missing CVE IDs or other important information due to timing of submission, etc. Developers should add information in a timely manner.}
% \ding{185} Provide useable reproduction environment. Our reproduction study in~\Cref{secreprofail} shows that the issues with environment deployment lead to many failures in reproduction. Therefore, developers should ideally be able to provide environments that support vulnerability reproduction. 

\subsubsection{For Platform Maintainers}
%For maintainers of PoC platforms, 
We recommend: 
\textbf{\textit{1)}} \textit{Ensure timely updates of PoC references for disclosed vulnerabilities.} 
As discussed in Section~\ref{secPoCCoverage}, many security platforms are no longer maintained, leading to the unavailability of PoC reporting data. 
Maintainers should periodically verify the validity of these references to ensure that the data remains accessible and up-to-date. This practice is crucial for maintaining the reliability and usefulness of PoC resources.
\textbf{\textit{2)}} \textit{Set a (well-designed) PoC template for submissions.} As discussed in~\Cref{secstudykey} and~\Cref{secreprofail}, employing a well-defined template for PoC submissions can significantly enhance the completeness and usability of PoC reports. Maintainers are encouraged to develop a PoC template that requires the submitter to include at least the key components we have identified. 
This practice will help ensure the consistency and completeness of the information provided, thereby facilitating PoC completeness.
\textbf{\textit{3)}} \textit{Implement a multi-faceted search strategy for PoCs.} 
Current platforms predominantly offer search capabilities limited to components like CVE IDs or PoC titles, which may not align well with the actual needs of users. 
To bridge this gap, platform maintainers should develop a multifaceted and user-friendly search strategy that includes ways for searching by key components of PoCs, and other relevant criteria. This approach will enhance both the efficiency and accuracy of searches, making it easier for users to quickly locate the specific PoCs they require. 
% We are in the process of building a multi-faced search for our website based on the already implemented aspect-level search capabilities (details in~\Cref{sec:data-application}). 

% The platform maintainers are expected to implement a multi-faceted PoC search strategy, including CVE ID-based or key component-based searches, etc. This approach can help improve the efficiency and accuracy of PoC retrieval, making it easier for users to find the relevant PoCs they need.

\subsubsection{For PoC Users}
We offer several practical guidance to enhance existing PoCs' usability: 
\textbf{\textit{1)}} \textit{Augment PoC completeness by combining multi-source information.}
As discussed in~\Cref{secstudykey} and~\Cref{sec:enhancerepro}, PoCs from various sources generally complement each other, filling gaps and providing a more comprehensive perspective for reproduction. 
To enhance reproducibility, users are advised to fuse information from multiple PoC reports about the same vulnerability. 
\textbf{\textit{2)}} \textit{LLM-assisted vulnerability reproduction.}
Our reproduction study in Section~\ref{sec:enhancerepro} demonstrated the effectiveness of LLMs for augmenting PoC reports and error diagnosis. 
Users can consider LLM-assisted reasoning for completing missing key components and handling reproduction errors, thereby improving reproducibility.  

\subsubsection{For PoC Researchers}
We discuss some future directions and encourage research to: 
\textbf{\textit{1)}} \textit{Establish missing links between PoCs and disclosed vulnerabilities.} 
A significant disconnect exists between many disclosed vulnerabilities (e.g., CVE entries) and corresponding PoCs, as highlighted in~\Cref{secPoCCoverage}, resulting in many PoCs being unusable.
% This gap results in many PoCs being unusable because it is unclear which disclosed vulnerabilities (e.g., CVE entries) they are meant to demonstrate, and vice versa. 
This limitation severely hampers the practical application and effectiveness of both the PoCs and the vulnerability documentation. 
We encourage future research to focus on establishing clear, bidirectional links between vulnerabilities and their respective PoCs, thereby enhancing the utility and effectiveness of both. 
\textbf{\textit{2)}} \textit{Augmentation of PoC data completeness.} Multi-source PoC information fusion and LLM-based key component reasoning have been proven to improve the success rate of reproduction in~\Cref{sec:enhancerepro}. It would substantially improve the completeness and usability of existing PoC reports if this process could be automated in future research. 
% \ding{184} \textit{PoC generation.} Section~\ref{secPoCCoverage} illustrates that the current lack of PoCs is very serious, and we call for PoC generation efforts that can cover more vulnerability types. 
\textbf{\textit{3)}} \textit{Enhance PoC generation efforts.} 
Despite numerous efforts in PoC generation in recent works, the current scarcity of PoCs across various vulnerability types is still severe (21.1\%), particularly for vulnerability types under CWE-125 and CWE-77 (details in Section~\ref{secPoCCoverage}). 
We advocate future research for targeted PoC generation designed to comprehensively cover these vulnerability types.
\dwj{\textbf{\textit{4)}} \textit{Prioritize PoC development for neglected vulnerability types.} Our measurement in~\Cref{secvultypes} revealed a significant difference in \poccover across different vulnerability types. 
Vulnerability types such as CWE-125 and CWE-77 have many disclosed vulnerability instances, underscoring their prevalence and importance. 
However, they have been overlooked in PoC generation and publication, with their \poccover being only about 10\%.  
Great attention is needed to develop PoCs for them to enhance understanding and improve defenses. }
%, thereby bridging existing gaps. 

\section{Threats to Validity and Limitations}
% \subsection{External Validity}~\label{external_threat}

% \subsection{Internal Validity}~\label{internal_threat}

% \revise{Why not miss this very important part!}
\textbf{\textit{1)}} The first threat concerns the comprehensiveness of data collection. 
Given that PoCs are distributed across numerous sources, it is challenging to collect them completely. 
We mitigate this by gathering as many PoC databases as possible through SLR and expanding our dataset using NVD references as discussed in~\Cref{sec:collectpoc}. 
%Moreover, collecting PoCs exclusively from CVEs is not practical for us, as there are thousands of sources linked through CVE references. 
%Existing methods such as~\cite{GitHubtr81:online} face severe limitations, as they rely on traversing CVE reference links and matching specific strings like ``PoC'' and ``Proof-of-Concept'', which can result in various misidentifications, as detailed on~\cite{notfromCVE}. 
Through our systematic collection process, we compiled the largest PoC dataset to date, which is sufficient for our study purpose. \textbf{\textit{2)}} The second threat is the impact of our method's inability to achieve complete accuracy in measuring the presence of key components. To mitigate this, we fine-tuned the BERT-NER model using 2,400 manually labeled PoC reports. As a result, our tool achieved an overall F1-score of 0.95. Although minor deviations remain, the general presence of key components can still be reliably summarized. \textbf{\textit{3)}} Another potential threat arises from bias in the manual reproduction of vulnerabilities. We mitigated this by enlisting \cs{ten participants} and dividing them into two groups.  The results from both groups were compared for consistency.
%Each group was tasked with reproducing the same 100 vulnerabilities. The results from both groups were compared for consistency. 
%In cases where the results differed, a discussion was held, and another attempt at reproduction was made until consistent results were achieved.

% }\revise{Confusing, may delete and talk this in threats to validity.

% Data collection

% Kait

% reproduce cross check/multiple

\section{Related Work}
%\noindent{\textbf{Existing research.}}
There is yet no large-scale empirical study directly focusing on PoC reports in the wild. 
Existing PoC researches generally focus on PoC generation~\cite{martin2008automatic, kieyzun2009automatic, alhuzali2016chainsaw, alhuzali2018navex, jan2017automatic, park2022fugio, brumley2008automatic, hu2015automatic, wu2018fuze, chen2020koobe, lee2021exprace, zhang2023automated, liang2024k} and application~\cite{bozorgi2010beyond,wang2018revery,suciu2022expected, dai2021facilitating, kwon2021octopocs, jiang2022evocatio, chen2023exploiting, yuan2021raproducer}. 

% \subsection{PoC Generation}

% \subsection{PoC Generation} 
Current works for PoC generation mainly focus on web applications~\cite{martin2008automatic, kieyzun2009automatic, alhuzali2016chainsaw, alhuzali2018navex, jan2017automatic, park2022fugio} and binary programs~\cite{brumley2008automatic, hu2015automatic, wu2018fuze, chen2020koobe, lee2021exprace, zhang2023automated, liang2024k}. Regarding those on web applications, some works focused on SQL Injection and XSS vulnerabilities~\cite{martin2008automatic, kieyzun2009automatic, alhuzali2016chainsaw, alhuzali2018navex}. 
% For instance, Martin et al.~\cite{martin2008automatic} and Ardilla et al.~\cite{kieyzun2009automatic} generated exploits using taint analysis and taint tracking methods, respectively. 
% Alhuzali et al.~\cite{alhuzali2016chainsaw} employed static analysis and symbolic execution to generate exploits with a complete HTTP request sequence and later introduced NAVEX\cite{alhuzali2018navex}, which was enhanced through dynamically coordinated execution.  
Similarly, several works are also conducted for other vulnerability types including XML~\cite{jan2017automatic} or PHP object injection~\cite{park2022fugio}. 
% For instance, Jan et al.~\cite{jan2017automatic} designed a search algorithm and fitness function for XML injection\cmt{check the type name} vulnerability exploit generation. 
% Park et al.~\cite{park2022fugio} combine program analysis methods and fuzz testing to generate vulnerability exploits for PHP object injection vulnerabilities. 
Regarding works on binary programs, Brumley et al.~\cite{brumley2008automatic} proved the possibility of generating exploits based on vulnerability patches. 
% Hu et al.~\cite{hu2015automatic} developed data stream splicing technology to exploit data vulnerabilities. 
Meanwhile, several works were conducted targeting specific vulnerability types, such as \textit{Use After Free}~\cite{wu2018fuze}, \textit{Out-Of-Bounds Write}~\cite{chen2020koobe}, \textit{Kernel Races}~\cite{lee2021exprace}, and \textit{Memory Leak}~\cite{liang2024k}. 
% many works were conducted especially for specific vulnerability types including CWE-89 (SQL Injection) and CWE-79 (Cross-site Scripting) ~\cite{martin2008automatic, kieyzun2009automatic, alhuzali2016chainsaw, alhuzali2018navex}. 
% For instance, Brumley et al.~\cite{brumley2008automatic} leverage Windows Update patches to traverse their paths for PoC generation. 
% SemFuzz~\cite{you2017semfuzz}, Chen et al.~\cite{chenefficient} and Oddfuzz~\cite{cao2023oddfuzz} use CVE description, compiled files of the tested program, and conditions based on language dynamic features as input to generate PoC seed for fuzzing. 
% Another approach in~\cite{yang20231dfuzz} addresses 1-day vulnerabilities, employing trailing call sequence (TCS) and static analysis to identify potential vulnerability locations, followed by fuzz testing to generate PoCs. 
% These research efforts represent a significant stride in automated PoC generation, yet they also highlight the challenges and limitations inherent in current methodologies. 
% The dependency on specific types of vulnerabilities or the quality of available information underscores the need for more versatile and robust approaches in the automated generation of PoCs within the security domain.

% \subsection{PoC Application} 
% \subsection{PoC Application}
Several research works have explored the application of PoC in downstream tasks~\cite{bozorgi2010beyond,wang2018revery,suciu2022expected, dai2021facilitating, kwon2021octopocs, jiang2022evocatio, chen2023exploiting, yuan2021raproducer}. 
% One of the primary uses of PoC is in predicting the exploitability of vulnerabilities. 
Many works predict the exploitability of vulnerabilities by leveraging PoCs as inputs~\cite{bozorgi2010beyond,wang2018revery,suciu2022expected}, in which they reflect the likelihood that functional exploits will be developed. 
% Specifically, Bozorgi et al.~\cite{bozorgi2010beyond} approaches exploitability assessment through the training of machine learning models, automating the evaluation process. Wang et al.~\cite{wang2018revery} primarily use PoC to identify crash paths of vulnerabilities, facilitating a series of subsequent calculations. Suciu et al.~\cite{suciu2022expected} estimate the likelihood of developing functional exploits for vulnerabilities using multiple indicators, including PoC. 
Meanwhile, several works focus on vulnerability validation, in which they use PoCs to aid in identifying the capabilities and scope of impact of a vulnerability~\cite{dai2021facilitating, kwon2021octopocs, jiang2022evocatio, chen2023exploiting}. 
% In addition, Yuan et al.~\cite{yuan2021raproducer} present RAProducer, which uses the execution paths within PoC code to address the challenge of reproducing data race-type bugs, which are notoriously difficult to reproduce. 
% There is no direct analysis of PoC code snippets or PoC reports at this time, while several works are focusing on the analysis of objects such as vulnerability reports, and PoCs are indirectly included in the discussion as a component of the objects they analyze. Researchers have delved into the composition of vulnerability reports, describing PoC's place and impact in~\cite{bettenburg2008makes} and~\cite{chaparro2017detecting}. Bettenburg et al.~\cite{bettenburg2008makes} claimed that vulnerabilities with code snippets (including PoC code) in their reports tend to be fixed faster and Chaparro et al.~\cite{chaparro2017detecting} explored missing information in vulnerability reports. 
% Currently, there is no direct analysis of PoC code or PoC reports. 
% Instead, several works focus on the analysis of broader objects, such as vulnerability reports, within which PoCs are indirectly analyzed as components. 

% \subsection{Studies on Bug Reports}
% \subsection{Studies on Bug Report}
Investigations into bug reports have provided valuable insights into vulnerability reproduction~\cite{mu2018understanding,chaparro2017detecting,bettenburg2008makes}. 
% They underscore the critical role of PoCs in understanding and reproducing vulnerabilities, indirectly motivating our research. 
Bettenburg et al.~\cite{bettenburg2008makes} reveal that PoCs are critical for both vulnerability verification and fixes. 
% Chaparro et al.~\cite{chaparro2017detecting} manually analyzed nearly 3k bug reports to examine the absence of critical information in vulnerability reports. They found that only 51.4\% of bug reports contained ``Steps to Reproduce'', highlighting their {low presence rate}. 
% Interestingly, our work reflects that about 80\% disclosed vulnerabilities lack the corresponding PoCs, which, to some extent, corroborates their findings. 
{Furthermore, Mu et al.~\cite{mu2018understanding} focused on \cs{understanding} the reproducibility of Linux memory-error vulnerabilities \cs{based on CVE reports including description information and external references} and found that reproducing vulnerabilities without direct PoCs is particularly challenging, thereby underscoring the indispensable role of PoCs.
\cs{They did not study PoC, and their reportswere  written specifically for vulnerability reproduction.}
In contrast, we focused on large-scale PoC reports \cs{from different dimensions}, including the PoC availability, completeness, and reproducibility, to explore the overall ecological status of PoCs.}

\section{Conclusion}
This paper presented a large-scale study on PoC reports in the wild, focusing on their usability status including availability, completeness, and reproducibility. 
\cs{Our findings revealed significant insights for this research field.} %including PoC availability, completeness, and challenges in reproducing vulnerabilities. 
\cs{Based on the outputs of our study, we further outlined actionable strategies for stakeholders to enhance the usability of PoCs and discussed new research directions to address the identified challenges.}%including automated PoC generation, standardized reporting formats, and PoC augmentation strategies. 
\bibliographystyle{plain}
\bibliography{ref}

%\newpage

\end{document}